\documentclass[12pt,preprint]{aastex}

\usepackage{times}

\catcode`\@=11
\newcommand{\gapprox}{\mathrel{\mathpalette\@versim>}}
\newcommand{\lapprox}{\mathrel{\mathpalette\@versim<}}
\newcommand{\propapprox}{\mathrel{\mathpalette\@versim\propto}}
\newcommand{\@versim}[2]
  {\lower3.1truept\vbox{\baselineskip0pt\lineskip0.5truept
\ialign{$\m@th#1\hfil##\hfil$\crcr#2\crcr\sim\crcr}}}
\catcode`\@=12

\shorttitle{TWO YOUNG TYPE IA SNRS WITH IRS AND RGS}
\shortauthors{WILLIAMS ET AL.}

\begin{document}

\title{Dusty Blastwaves of Two Young LMC Supernova Remnants:
Constraints on Postshock Compression}

\author{Brian J. Williams,\altaffilmark{1}
Kazimierz J. Borkowski,\altaffilmark{1}
Stephen P. Reynolds,\altaffilmark{1}
Parviz Ghavamian,\altaffilmark{2}
John C. Raymond,\altaffilmark{3}
Knox S. Long,\altaffilmark{2}
William P. Blair,\altaffilmark{4}
Ravi Sankrit,\altaffilmark{5}
R. Chris Smith,\altaffilmark{6}
Sean Points,\altaffilmark{6}
P. Frank Winkler\altaffilmark{7}
\& Sean P. Hendrick\altaffilmark{8}
}

\altaffiltext{1}{Physics Dept., North Carolina State U.,
    Raleigh, NC 27695-8202; bjwilli2@ncsu.edu}
\altaffiltext{2}{STScI, 3700 San Martin Dr., Baltimore, MD 21218}
\altaffiltext{3}{Harvard-Smithsonian Center for Astrophysics, 60 Garden
    Street, Cambridge, MA 02138}
\altaffiltext{4}{Dept. of Physics and Astronomy, Johns Hopkins University, 
    3400 N. Charles St., Baltimore, MD 21218-2686}
\altaffiltext{5}{SOFIA/USRA}
\altaffiltext{6}{CTIO, Cailla 603, La Serena, Chile}
\altaffiltext{7}{Dept. of Physics, Middlebury College, Middlebury, VT 
    05753}
\altaffiltext{8}{Physics Dept., Millersville U., PO Box 1002, Millersville, 
    PA 17551}

\begin{abstract}

We present results from mid-IR spectroscopic observations of two young
supernova remnants (SNRs) in the Large Magellanic Cloud (LMC) done
with the {\it Spitzer Space Telescope}. We imaged SNRs B0509-67.5 and
B0519-69.0 with {\it Spitzer} in 2005, and follow-up spectroscopy
presented here confirms the presence of warm, shock heated dust, with
no lines present in the spectrum. We use model fits to {\it Spitzer}
IRS data to estimate the density of the postshock gas. Both remnants
show asymmetries in the infrared images, and we interpret bright spots
as places where the forward shock is running into material that is
several times denser than elsewhere. The densities we infer for these
objects depend on the grain composition assumed, and we explore the
effects of differing grain porosity on the model fits. We also analyze
archival {\it XMM-Newton} RGS spectroscopic data, where both SNRs show
strong lines of both Fe and Si, coming from ejecta, as well as strong
O lines, which may come from ejecta or shocked ambient medium. We use
model fits to IRS spectra to predict X-ray O line strengths for
various grain models and values of the shock compression ratio. For
0509-67.5, we find that compact (solid) grain models require nearly
all O lines in X-ray spectra to originate in reverse-shocked
ejecta. Porous dust grains would lower the strength of ejecta lines
relative to those arising in the shocked ambient medium. In 0519-69.0,
we find significant evidence for a higher than standard compression
ratio of 12, implying efficient cosmic-ray acceleration by the blast
wave. A compact grain model is favored over porous grain models. We
find that the dust-to-gas mass ratio of the ambient medium is
significantly lower than what is expected in the ISM.

\end{abstract}


\section{Introduction}

Supernova remnants (SNRs) provide a laboratory to study various
aspects of interstellar medium (ISM) evolution across the whole
electromagnetic spectrum. With the advent of high spatial resolution
telescopes in both the X-ray and infrared (IR) regimes, it is possible
to probe the interaction of the rapidly moving shock wave with the
dust and gas of the surrounding ambient medium and to study the ejecta
products of the SN itself. The expanding shockwave sweeps up and heats
gas to $10^{6}-10^{9}$ K, causing it to shine brightly in X-rays. Dust
grains embedded in the hot, shocked plasma are collisionally heated,
causing them to radiate at IR wavelengths, and slowly destroying them
in the process via sputtering. Because the physical processes behind
X-ray and IR emission are related, a combined approach to studying
SNRs using both energy ranges can reveal more information than either
could on its own.

To better characterize IR emission from SNRs, we conducted an imaging
survey with the {\it Spitzer Space Telescope} of $\sim 40$ known SNRs
in the Large and Small Magellanic Clouds. The Clouds were chosen
because of their known distance and relatively low Galactic IR
background. Subsequently, we obtained IR spectra of several of the
SNRs in the LMC using the Infrared Spectrograph (IRS) on {\it
Spitzer}. We report here on IRS observations of two of these, SNRs
B0509-67.5 (hereafter 0509) and B0519-69.0 (hereafter 0519). Both are
remnants of thermonuclear SNe \citep{smith91,hughes95} and have fast,
non-radiative shocks (several thousand km s$^{-1}$)
\citep{tuohy82,ghavamian07}.  There is no evidence in either for
slower, radiative shocks. They are both located in the LMC, at the
known distance of $\sim 50$ kpc. In addition, both have ages
determined from light echoes \citep{rest05}, with 0509 being 400 $\pm
120$ and 0519 being 600 $\pm 200$ years old. They are nearly identical
in size, having an angular diameter of $\sim 30$'', which corresponds
to a physical diameter of $\sim 7.3$ pc.

In \citet{borkowski06} hereafter Paper I, we used 24 and 70 $\mu$m
imaging detections of both SNRs to put limits on the post-shock
density and the amount of dust destruction that has taken place behind
the shock front, as well as put a limit on the dust-to-gas mass ratio
in the ambient medium, which we found to be a factor of several times
lower than the standard value for the LMC of $\sim 0.25$\%
(Weingartner \& Draine 2001, hereafter WD01). These limits, however,
were based on only one IR detection, at 24 $\mu$m, with upper limits
placed on the 70 $\mu$m detection. Both remnants were detected with
{\it Akari} \citep{seok08} at 15 and 24 $\mu$m (0519 was also detected
at 11 $\mu$m). Seok et al. applied single-temperature dust models to
{\it Akari} data, deriving warm dust masses of 8.7 $\times 10^{-5}$
$M_\odot$ and $3.6 \times 10^{-4}$ $M_\odot$ in 0509 and 0519,
respectively. With full spectroscopic data, we can place much more
stringent constraints on the dust destruction and dust-to-gas mass
ratio in the ISM. We also explore alternative dust models, such as
porous and composite grains.

Additionally, we examine archival data from the Reflection Grating
Spectrometer (RGS) onboard the {\it XMM-Newton} X-ray observatory. The
high spectral resolution of RGS allows us to measure the strength of
lines in X-ray spectra. We can use post-shock densities derived from
IR fits to predict the strength of lines arising from shocked ambient
medium, and can thus derive relative strengths of ejecta contributions
to oxygen lines. Doing this requires knowledge of the shock
compression ratio, $r$, defined as $n_{H}/n_{0}$ (where $n_{H}$ and
$n_{0}$ are postshock and preshock hydrogen number densities,
respectively), which for a standard strong shock (Mach number $\gg$ 1)
is 4. However, shock dynamics will be modified \citep{jones91} if the
shock is efficient at accelerating cosmic rays, and $r$ can be
increased by a factor of several. We use several representative values
of $r$ in modeling X-ray line strengths.

\section{Observations and Data Reduction}
\label{obs}

We mapped both objects using the long wavelength (14-38 $\mu$m),
low-resolution ($\Delta\lambda/\lambda$ 64-128) (LL) spectrometer on
{\it Spitzer's} IRS (Program ID 40604). For 0509, we stepped across
the remnant in seven LL slit pointings, stepping perpendicularly
5.1$''$ each time. This step size is half the slit width, and is also
the size of a pixel on the LL spectrograph. We then shifted the slit
positions by 56$''$ in the parallel direction and stepped across the
remnant again, ensuring redundancy for the entire object. This process
was repeated for each of the two orders of the spectrograph. Each
pointing consisted of two 120-second cycles, for a total observing
time of 6720 s.

For 0519, the process was identical in terms of number of positions
and step sizes in between mappings, but each pointing consisted of
four 30-second cycles, for a total observing time of 3360 s. The
difference was due to the fact that 0519 is several times brighter in
the wavelength range of interest, and we did not want to saturate the
detectors. 

To estimate the uncertainties in the data, we fit a second-order
polynomial to the line-free wavelength regions ($\sim 21-32$ $\mu$m)
in the spectra where the signal-to-noise was adequate for
modeling. The standard deviations of the data points in these regions
were used for $\chi^{2}$ fitting of dust emission models (see
Section~\ref{errors}). We obtained $\sigma$ values of 0.830 and 2.53
mJy for 0509 and 0519, respectively.

The spectra were processed at the {\it Spitzer Science Center} using
version 17.1 of the IRS pipeline. We ran a clipping algorithm on the
spectra which removes both hot and cold pixels that are more than
3-$\sigma$ away from the average of the surrounding pixels. To extract
the spectra, we used SPICE, the {\it Spitzer} Custom Extraction
tool. Once all spectra were extracted, we stacked spectra from the
same spatial location, improving the signal-to-noise ratio of the
sources. For background subtraction, we use the off source slit
positions that come when one of the two slit orders is on the
source. In the end, we have seven (overlapping) background subtracted
spectra for each remnant. As we show in Section~\ref{results}, this
allows us to do spatially resolved spectroscopy, despite the fact that
the remnants are only $\sim 30''$ in diameter.

We processed archival {\it XMM-Newton} RGS data from the XMM Science
Archive with version 8.0 of the Science Analysis Subsystem (SAS)
software for XMM. RGS provides high-resolution ($\Delta\lambda$ $\sim
0.04$ \AA) X-Ray spectroscopy in the wavelength range from 5-38
\AA. B0509-67.5 was observed on 4 July 2000 (Obs. ID 0111130201, PI
M. Watson) for 36 ksec. Of these 36 ksec, only the last $\sim 2500$ s
were affected by flaring. We discarded these and use the remaining
33.5 ksec for our analysis. We use the spectra from both RGS 1 and 2,
and bin the data using the FTOOL {\it grppha} to a minimum of 25
counts per bin. Both RGS orders contained $\sim 8000$ counts after
time-filtering. B0519-69.0 was observed on 17 September 2001 (Obs. ID
0113000501, PI J. Kaastra) for 48.4 ksec. After time filtering, we
obtained $\sim 35$ ksec of useful data, with $\sim 12000$ counts in
both RGS 1 and 2. We also grouped these by a minimum of 25 counts per
bin. Since RGS is a slitless spectrometer, spatial information is
degraded for extended sources, and the spectrum is smeared by the
image of the source. In order to model RGS spectra, the response files
generated by SAS must be convolved with a high-resolution X-ray
image. For that purpose, we used archival broadband Chandra images
from 2001 (Obs. ID 776, PI J. Hughes), along with the FTOOL {\it
rgsrmfsmooth}, to produce new response matrices.

\section{Results}
\label{results}

\subsection{0509}

In the top left panel of Figure~\ref{24um}, we show the MIPS 24 $\mu$m
image of 0509, with overlays as described in the caption. Immediately
obvious is the large asymmetry in 0509, where the remnant brightens
from the faint NE hemisphere to the brighter SW. Dividing the remnant
in half and measuring the 24 $\mu$m fluxes from each half, we find a
flux ratio of five between these two regions. At X-ray wavelengths, we
measure a much more modest ratio of 1.5 from archival broadband {\it
Chandra} images. Using IRS, we separated the remnant into 7 overlapping
regions from which spectra were extracted. For our analysis, we chose
two regions (shown on Figure~\ref{24um}) that do not overlap spatially
and provide adequate signal to noise spectra. These regions roughly
correspond to the bright and faint halves of the remnant, and the
spectra are shown in Figure~\ref{faintvsbright}. In
Figure~\ref{0509faint} and Figure~\ref{0509bright}, we show both of
these spectra separately, with models overlaid (models are discussed
in Section~\ref{modeling}). The ratio of integrated fluxes from 14-35
$\mu$m for the two regions is $\sim 2.5$. This differs from the factor
of 5 measured from the photometric images because the slits are not
exactly aligned with the bright and faint halves of the remnant, and
because of the differences in the MIPS and IRS bandpasses.

A few obvious features immediately stand out about both
spectra. First, there is no line emission at all seen in either
spectrum. This is not unexpected, since the shocks in 0509 are some of
the fastest SNR shocks known, at $> 5000$ km s$^{-1}$
\citep{ghavamian07}, and there is no optical evidence for radiative
shocks. Second, though both spectra show continuum from warm dust,
there are obvious differences in the spectra. An inflection around 18
$\mu$m can be seen in the spectrum from the bright half, while this
feature is not as clear in the faint spectrum. We attribute this
feature to the Si-O-Si bending mode in amorphous silicate dust. In
Section~\ref{disc}, we will explore reasons for the different spectral
shapes and the differences in brightness between the two halves of the
remnant.

\subsection{0519}

In the bottom left panel of Figure~\ref{24um}, we show the 24 $\mu$m
image of 0519, which does not show the large scale asymmetries that
0509 does, although three bright knots can be seen in the image. These
same three knots can be identified in both H$\alpha$ \citep{smith91}
and X-ray images, where {\it Chandra} broadband data shows that the
three knots are prominent in the 0.3-0.7 keV band
\citep{kosenko10}. In Paper I, we measured the total flux from the
three knots added together and found that collectively they represent
only about 20\% of the flux. Nevertheless, spectra extracted from
slits that contain a bright knot do show differences in continuum
slope from those extracted where no knots are present. We discuss our
interpretation of these knots in Section~\ref{disc}. As with 0509,
there are no lines seen in the IRS data (see Figure~\ref{0519irs}), as
shock speeds are also quite high in this object.

\section{IRS Fits}
\label{modeling}

We now turn our attention to modeling the emission seen in IRS with
numerical models of collisionally heated dust grains. We follow a
procedure identical to that followed in our previous work on these
objects; see Paper I. The heating rate of a dust grain immersed in a
hot plasma depends on both the gas density (collision rate) and gas
temperature (energy per collision). In general, the electron
temperature of a shocked plasma can be determined from fits to X-ray
spectra. In the case of these objects, plane-shock models without
collisionless heating of electrons at the shock front yield the
temperature. Dust heating models do depend on this temperature, but
the dependency is not large \citep{dwek87}. Gas density is a much more
sensitive parameter for dust heating models, and is determined by the
temperature of the grains. We regard postshock gas density as a free
parameter in our model, and fine tune the density to match the
observed IR spectra.

We take into account sputtering by ions, which destroys small grains
and sputters material off large grains. We use a plane-shock model
which superimposes regions of increasing shock age (or sputtering
timescale) $\tau_p=\int_0^t n_p dt$, where $n_{p}$ is post-shock
proton density, while keeping temperature behind the shock
constant. Inputs to the model are an assumed pre-shock grain-size
distribution, grain type and abundance, proton and electron density,
ion and electron temperature. The code calculates the heating and
sputtering for grains from 1 nm to 1 $\mu$m, producing a grain
temperature and spectrum for each grain size. The spectra are then
added in proper proportions according to the post-sputtering size
distribution to produce a model spectrum that can be compared with
observations. We use sputtering rates from \citet{nozawa06} with
enhancements for small grains from \citet{jurac98}. We assume that
sputtering yields are proportional to the amount of energy deposited
by incoming particles, accounting for partial transparency of grains
to protons and alpha particles at high energies
\citep{serradiazcano08}, particularly relevant for 0509 and 0519. We
assume that grains are compact spheres, but we also report results and
implications if grains are porous (i.e. contain a nonzero volume
filling fraction of vacuum).

In general, our method is to fix as many of the input parameters as
possible, based on what is known from other observations. For
instance, \citet{rest05} used optical light echoes to constrain the
ages of both remnants, yielding ages of $\sim 400$ yrs. for 0509 and
$\sim 600$ yrs. for 0519. Shock velocities are known approximately for
both objects from measurements of ion temperatures
\citep{tuohy82,ghavamian07}. In 0509, the standard shock jump
conditions for a 7000 km s$^{-1}$ shock, the assumed speed for the NE
half of the remnant \citep{ghavamian07} with no ion-electron
equilibration would produce proton temperatures of $\sim 115$
keV. Recently, \citet{helder10} used VLT spectroscopy to detect a
broad H$\alpha$ component from the rim of 0509, determining the shock
speed to be $\sim 6000$ km s$^{-1}$ in the NE. They give several
possibilities for the proton temperature in the NE, and note that
energy from escaping cosmic-rays limits the post-shock proton
temperature to a value of $<$ 0.85 of that expected from the strong
shock jump conditions. Adopting this value of 0.85, along with a shock
speed of 6000 km s$^{-1}$ and no electron heating gives a post-shock
proton temperature of $<$70 keV. We use this value as an input to dust
heating models, except where noted otherwise (see
Section~\ref{iontemp}). This upper limit on proton temperature makes
any fit to the density for 0509 a lower limit, and we discuss the
effects of changing the proton temperature in
Section~\ref{iontemp}. 

The shock speed in 0519 is a bit more uncertain. \citet{ghavamian07}
quote a range of 2600-4500 km s$^{-1}$ between the limits of minimal
and full ion-electron equilibration. \citet{kosenko10} measured line
widths of X-ray iron lines seen in {\it XMM-Newton} RGS spectra to
obtain a velocity of 2927 km s$^{-1}$. We adopt a value of 3000 km
s$^{-1}$ for this work and use the standard shock jump proton
temperature assuming no electron-ion equilibration, which is 21
keV. The actual proton temperature may differ from this estimate, but
the fitting of dust spectra in the IR is only weakly dependent on the
proton temperature (see Section~\ref{iontemp}).

The overall normalization of the model to the data (combined with the
known distance to the LMC) provides the mass in dust. Since our models
calculate the amount of sputtering that takes place in the shock, we
can then determine the amount of dust present in the pre-shock
undisturbed ISM that has been encountered by the blast wave. While we
are sensitive only to ``warm'' dust, and could in principle be
underestimating the mass in dust if a large percentage of it is too
cold to radiate at IRS wavelengths, it is difficult to imagine a
scenario where any significant amount of dust goes unheated by such
hot gas behind the high-velocity shocks present in both remnants,
particularly given the upper limits at 70 $\mu$m reported in
\citet{borkowski06} (and the fact that there are no radiative shocks
observed anywhere in these remnants).

\subsection{Model Uncertainties}
\label{errors}

The only fitted parameters in the model are the post-shock proton
number density, which affects the shape of the model spectrum, and the
radiating dust mass, which affects the overall normalization. All
other variable parameters (e.g. electron density, sputtering
timescale) depend directly on the density. We use an optimization
algorithm to minimize the value of $\chi^{2}$ in the regions of the
spectra with high signal-to-noise (for both remnants, $\sim 21-32$
$\mu$m) to find the best-fitting value of the density. We then
calculate upper and lower limits on the fitted density as the 90\%
confidence intervals in $\chi^{2}$ space (where confidence intervals
correspond to $\chi^{2}_{min}$ + 2.71), using the standard deviations
listed in Section 2. The errors on both the density and dust mass
present are of order 25\% for 0509 and 10\% for 0519. Larger errors in
0509 are due to a poorer signal-to-noise ratio in the spectrum. We
assume that the plasma temperature is fixed at the values listed
above, but again note that the dependency on proton and electron
temperature is small (see Section~\ref{iontemp}).

There are also several sources of uncertainty in modeling emission
from warm dust grains that are hard to quantify statistically. First
and foremost, we are limited by both the spatial resolution of {\it
Spitzer} for such small SNRs and the low signal-to-noise ratio of the
spectra. We use a plane-shock model with constant plasma temperature
to model dust emission, which is only an approximation to a spherical
blast wave. Estimates of swept gas mass assume constant ambient
density ahead of the shock. Sputtering rates for such high temperature
protons and alpha particles are probably only accurate to within a
factor of 2 (but see Appendix A), and we use the simple approximation
that the sputtering rate varies according to the amount of energy
deposited into a grain of given size and composition. Lastly, but
possibly most importantly, we do not know the degree of porosity and
mixing of various grain types.

\subsection{IR Morphologies}

Although we have full spectral mapping of both objects, the 10.5$''$
width of the IRS slit severely limits the amount of spatially resolved
spectroscopy we can do on remnants that are only 30$''$ in diameter,
as both 0509 and 0519 are. Despite this, we do have several disjoint
slit positions which isolate spectra from the most prominent features
in the images. Because dust radiates as a modified blackbody spectrum,
emission from a small amount of warmer dust can overwhelm that from a
larger amount of colder dust, particularly at the wavelengths of
interest here. We show in Figure~\ref{0509faint} and
Figure~\ref{0509bright} spectra extracted from two regions of 0509,
which we label the ``faint'' and ``bright'' portions of the
remnant. Fits to these regions require a density contrast of $\sim 4$
in the post-shock gas, with the higher density required in the
``bright'' region (higher densities means hotter dust, hence more
short wavelength emission). Although it might be possible that a
density gradient of this order actually exists in the ambient medium
surrounding 0509 (the angular size of 30$''$ in the LMC corresponds to
a linear diameter of about 7 pc), the implications from this model
lead to several scenarios which are unlikely and require an appeal to
special circumstances.

\begin{figure}
\includegraphics[width=12cm]{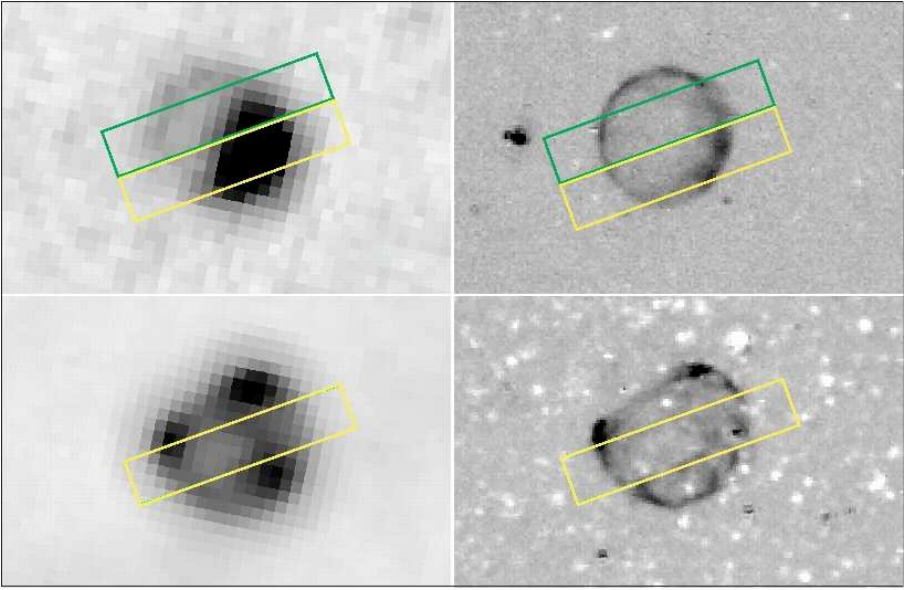}
\caption{Top left: MIPS 24 $\mu$m image of SNR B0509-67.5, overlaid
with regions of spectral extraction as described in text, where green
region is ``faint'' region and yellow marks ``bright'' region. Top
right: Star-subtracted H$\alpha$ image of 0509. Bottom left: MIPS 24
$\mu$m image of SNR B0519-69.0, with extraction slit shown in
yellow. Bottom right: Star-subtracted H$\alpha$ image of
0519. Spectral extraction regions in both remnants are approximately
10.5$''$ $\times$ 45$''$. FWHM of MIPS 24 $\mu$m PSF is approximately
7$''$. H$\alpha$ images obtained with the 4-m Blanco telescope at
NOAO/CTIO. In all images, north is up and east is to the left.
\label{24um}
}
\end{figure}

\begin{figure}
\includegraphics[width=17cm]{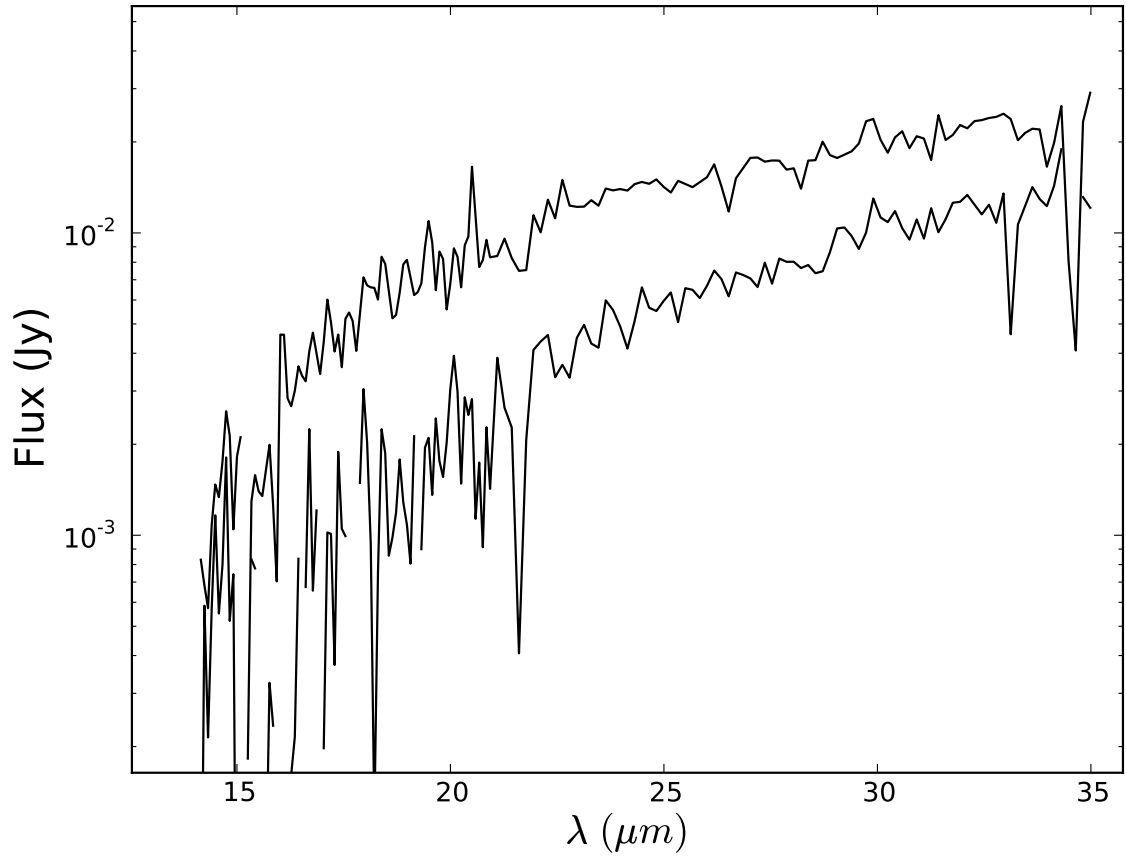}
\caption{Top curve: {\it Spitzer} IRS spectrum of ``bright'' region of
  0509. Bottom curve: spectrum of ``faint'' region. Note the
  differences in the slope of the spectra, in addition to the overall
  intensity.
\label{faintvsbright}
}
\end{figure}

\begin{figure}
\includegraphics[width=16cm]{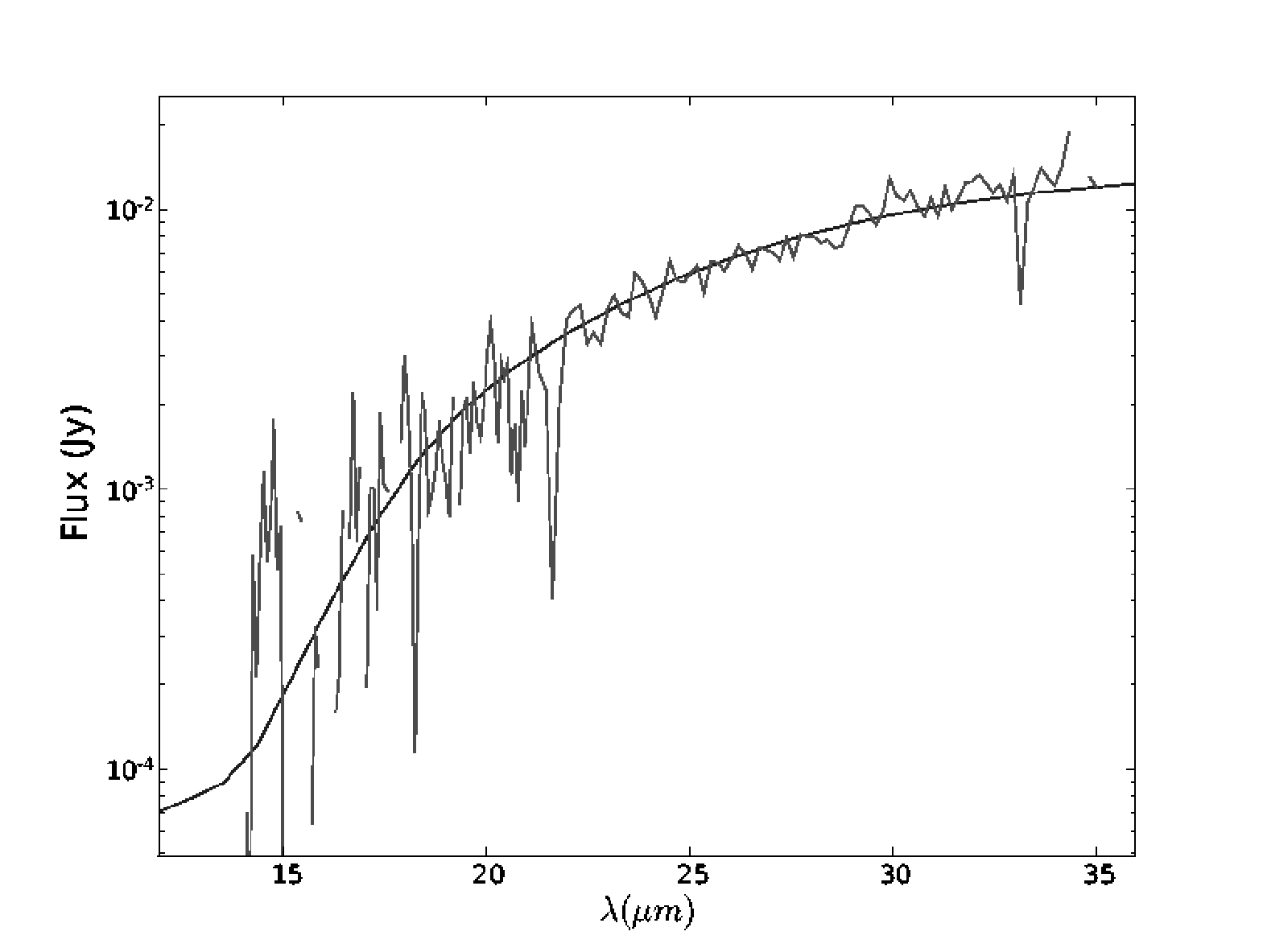}
\caption{14-35 $\mu$m IRS spectrum of the ``faint'' region of 0509, overlaid with model fit.
\label{0509faint}
}
\end{figure}

\begin{figure}
\includegraphics[width=16cm]{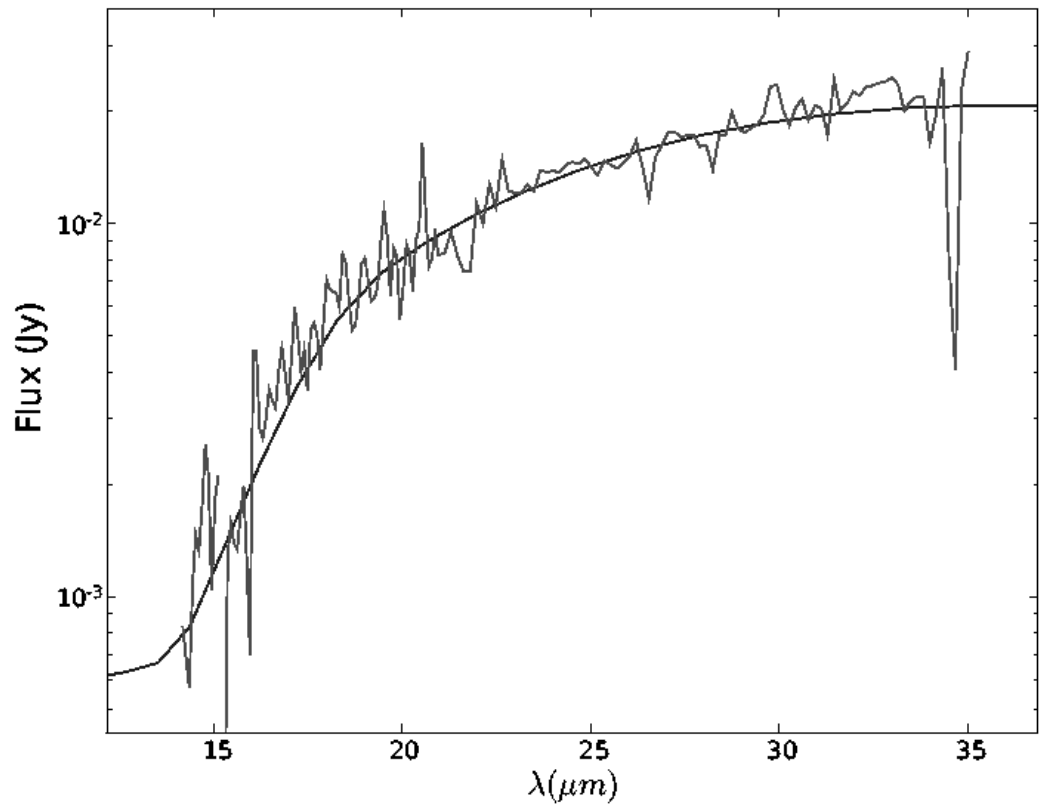}
\caption{14-35 $\mu$m IRS spectrum of the ``bright'' region of 0509, overlaid with model fit.
\label{0509bright}
}
\end{figure}

In comparing the MIPS image with images of the remnant at other
wavelengths, one can clearly see an enhancement in the {\it Chandra}
broadband image \citep{warren04} on the southwest side of the
remnant. This enhancement is mostly present in the energy range
containing Fe L-shell emission lines. The H$\alpha$ image, shown in
Figure~\ref{24um}, shows a uniform periphery around the remnant, with
the exception of a brightness enhancement in the SW, relatively well
confined to a small region. To compare the images, we convolved the
H$\alpha$ image to the resolution of the 24 $\mu$m image using the
MIPS 24 $\mu$m PSF. The result is shown in
Figure~\ref{halpha_conv}. The images are morphologically similar, and
from these comparisons we conclude that there is not an overall NE-SW
density gradient in the ISM. Rather, the ISM is mostly uniform except
for the SW, where the remnant is running into a localized region of
higher density. To obtain the conditions for the whole of the remnant,
we fit a model to the ``faint'' region, freezing all parameters to
those reported in Table 1 and allowing post-shock density to vary
(which also causes the sputtering timescale to vary). If we assume a
standard LMC dust model \citep{weingartner01}, we get a post-shock
density of $n_{H}$ = 0.59 cm$^{-3}$. The reduced $\chi^{2}$ value for
this model was 1.04. Values in Table~\ref{resultstable0509}
correspond to this ``faint'' region.

We can use IRS to estimate the spatial extent of the bright
region. First, by looking at the 2-D spectra from the LL slit located
directly on top of the brightest region in the MIPS 24 $\mu$m image,
we fit a gaussian to the spatial profile (indicated by an arrow in
Figure~\ref{24um_7points}) of the emission between 16 and 17 $\mu$m,
and found a FWHM of the brightness profile of 12.7''. This wavelength
is the shortest at which we could achieve a good signal-to-noise ratio
for this region. The {\it Spitzer} resolution at 16 $\mu$m is about
4.75''. In the direction perpendicular to the slit positions, we had 7
slit pointings, each separated by 5.1''. Extracting spectra from each
of these (overlapping) positions, subtracting the background, and
integrating over the IRS bandpass gave us 7 measurements of the
brightness as a function of position. The locations from which these
spectra were extracted are shown as diamonds in
Figure~\ref{24um_7points}, and the resulting normalized intensity
function is shown in Figure~\ref{intensitymap}. The brightness peaks
at position 3, the position we used to examine 2-D spectra, as
described above, and falls off relatively quickly on both sides away
from that position.

Our assumption that the bright region of 0509 is the result of a more
localized enhancement in density implies that the spectrum extracted
from that region is then a composite of emission from the uniform
parts of the remnant and from the small, denser region. {\it Spitzer}
does not have the resolution required to isolate this dense region,
but we can make a crude isolation by assuming that the two slit
positions in Figure~\ref{24um} cover an equal surface area on the
remnant, and subtracting the faint spectrum from the bright. The
residual spectrum is of less than ideal signal-to-noise. Nevertheless,
a fit to this spectrum implies a gas density of $\sim 6.4$ cm$^{-3}$,
or about an order of magnitude higher than the rest of the remnant. A
small region of hot dust outshines a more massive region of cooler
dust at 24 $\mu$m, which would explain the factor of 5 ratio in the
flux between the two sides. X-rays from this object are dominated by
ejecta, not swept-up ISM, and there are several factors to consider
beyond density when considering the flux of H$\alpha$ coming from a
region. Higher resolution observations at all wavelengths will shine
more light on this issue.

\begin{figure}
\includegraphics[width=16cm]{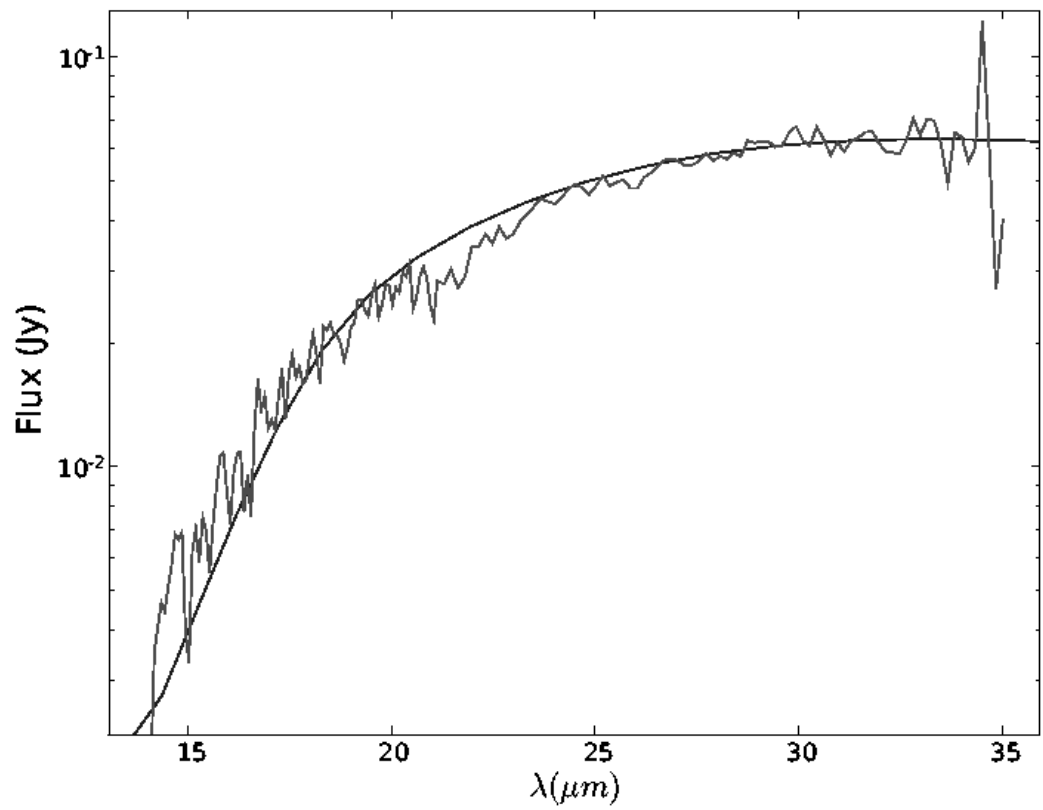}
\caption{14-35 $\mu$m IRS spectrum of 0519, extracted from a slit
placed across the middle of the remnant, free of emission from the
bright knots seen in the 24 $\mu$m image; with model overlaid.
\label{0519irs}
}
\end{figure}

\begin{figure}
\includegraphics[width=15cm]{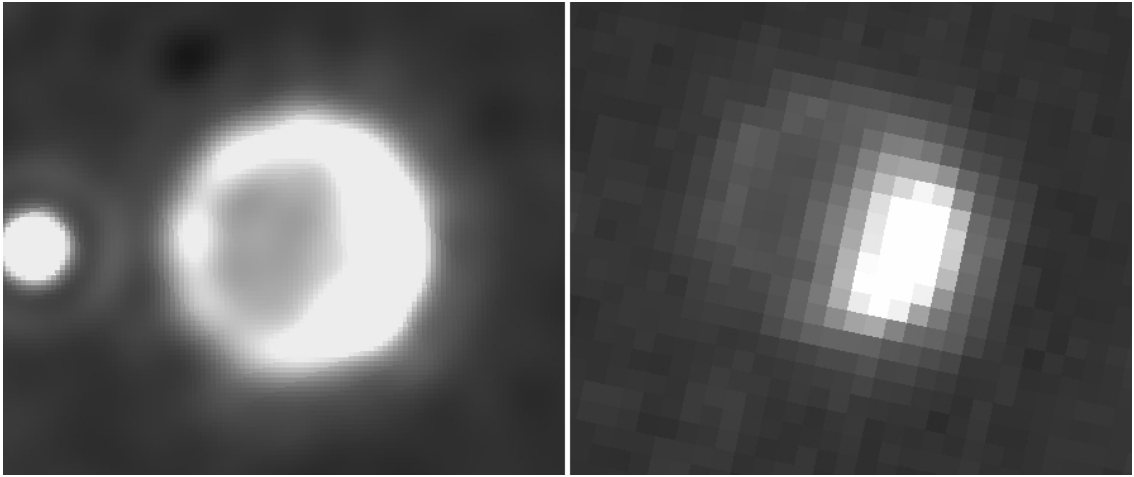}
\caption{Left: H$\alpha$ image of B0509-67.5 from Figure~\ref{24um},
convolved with the MIPS 24 $\mu$m PSF. Right: MIPS 24 $\mu$m image, at
native resolution.
\label{halpha_conv}
}
\end{figure}

\begin{figure}
\includegraphics[width=15cm]{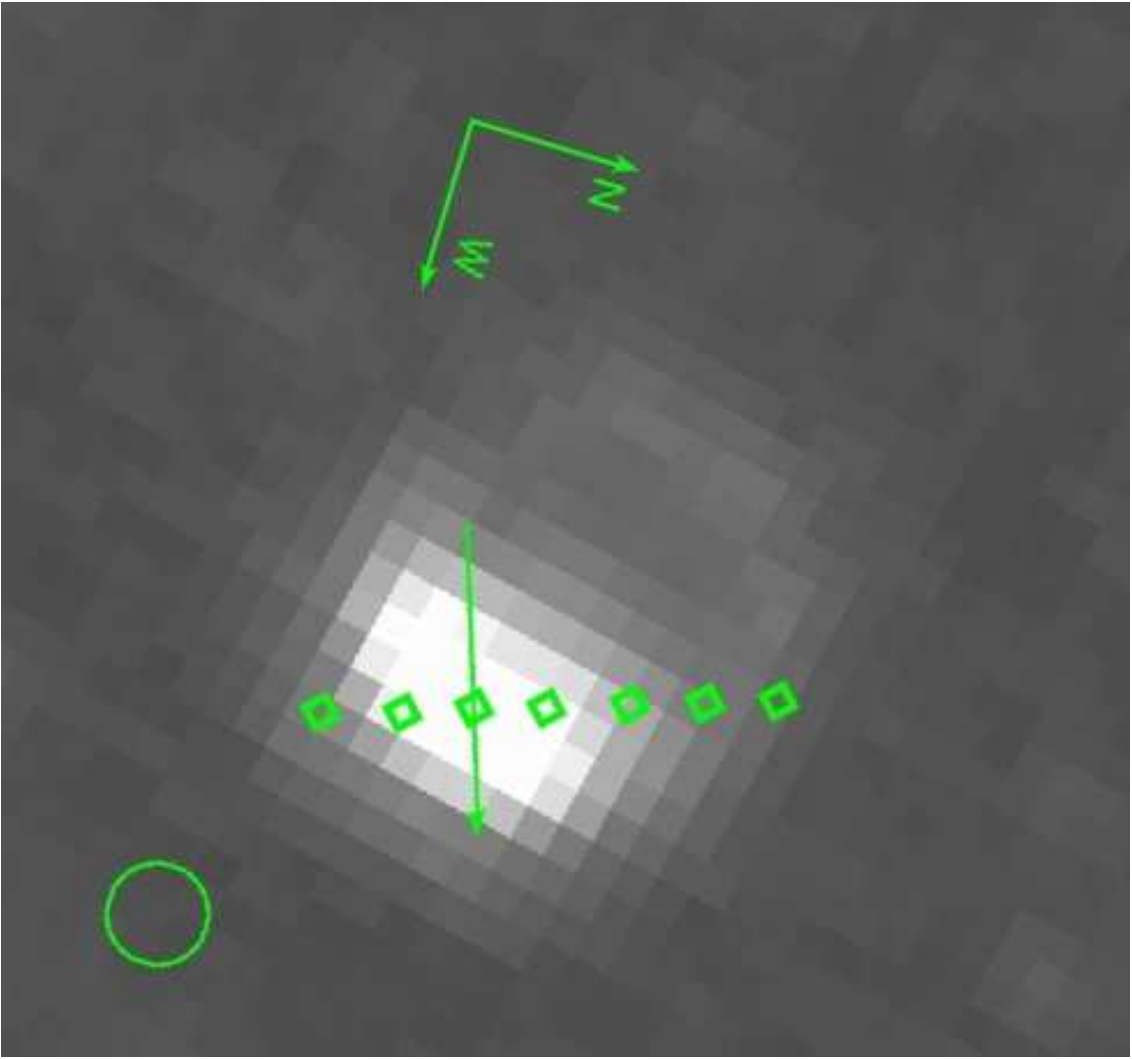}
\caption{24 $\mu$m image of 0509. Arrow shows both direction of
slit placement and length of inferred brightness enhancement in that
spatial direction. Diamonds are extraction centers for the parallel
slit positions extracted in the text (note that arrow coincides with
position 3). Intensity function plotted in Figure~\ref{intensitymap}
corresponds to diamonds from left-to-right in this image. Green circle
is MIPS 24$\mu$m PSF.
\label{24um_7points}
}
\end{figure}

\begin{figure}
\includegraphics[width=15cm]{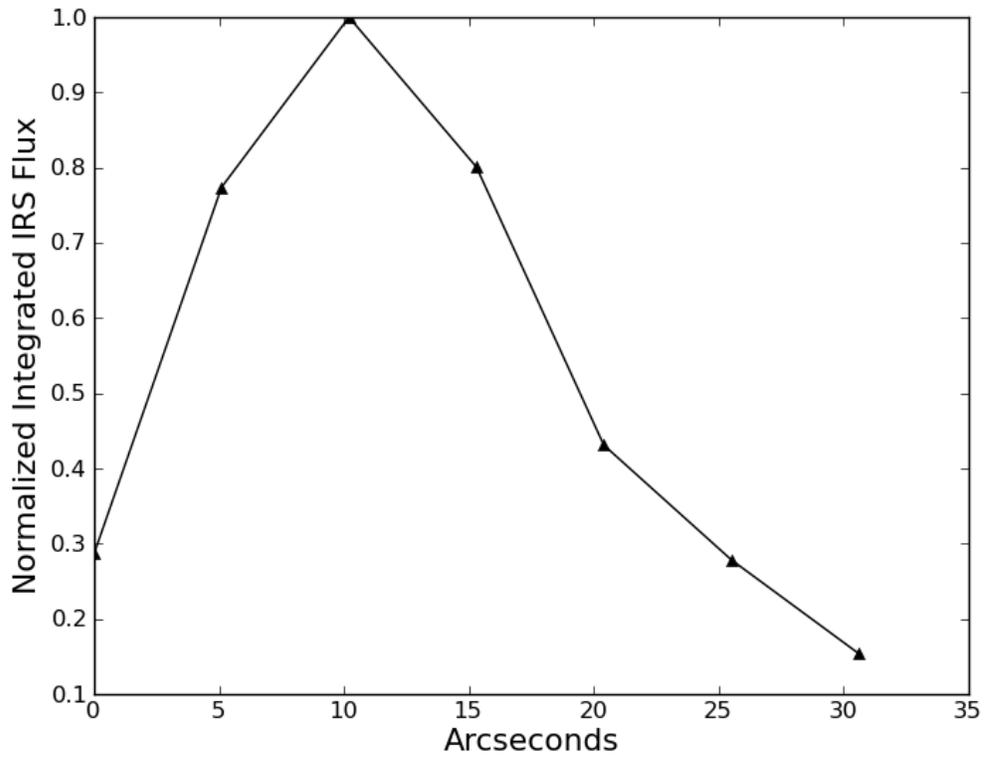}
\caption{Intensity of background-subtracted integrated IRS flux as a
function of distance from the left-most to the right-most diamond point
in Figure~\ref{24um_7points}.
\label{intensitymap}
}
\end{figure}

For 0519 we have a similar scenario, except that instead of one bright
region, we have three bright knots. As noted previously, these knots
correspond spatially with knots seen in both X-rays and H$\alpha$ (see
Figure~\ref{24um}). We adopt an identical strategy here, isolating
spectra from slit positions that do not overlap with one of the bright
knots. As it turns out, there is only one slit position in this object
that is nearly completely free of emission from a knot, a slit
position that goes directly across the middle of the remnant. The
spectrum from this region is shown in Figure~\ref{0519irs}. We assume
that the conditions within this slit position are indicative of the
remnant as a whole. Using parameters found in Table 1, we obtain a
post-shock density of $n_{H}$ = 6.2 cm$^{-3}$ (reduced $\chi^{2}$ =
1.13). We do not have adequate signal-to-noise to isolate the bright
knots individually. However, we approximate the density required by
noting that a fit to the spectrum extracted from the entire slit
position that contains the brightest of the three knots (the
northernmost knot) requires a density that is higher by a factor of
$\sim 2$. Given that this spectrum contains emission from both the
uniform parts of the remnant and the bright knot, it is likely that
the density in the knot itself is perhaps a factor of $\sim 3$ higher
that the more uniform parts of the remnant, which is roughly
consistent with our results from Paper I.

\section{Dust Grain Porosity}

While it is typically assumed in dust grain models that grains are
solid bodies composed of various materials, this is likely not a
physically valid model (see \citet{shen08}, and references
therein). It is possible to relax the assumption that grains are
compact spheres with a filling fraction of unity. We have updated dust
models to include the porosity of dust grains (where porosity, $\cal
P$, is the fraction of the grain volume that is composed of vacuum,
instead of grain material), as well as allowing for grains to be
composite in nature, i.e. a grain made up of silicates, graphite,
amorphous carbon, and vacuum. The details of this model are discussed
in \citet{williams10}. We summarize here the important
results. Perhaps the most significant effect of introducing porosity
to grains is that the grain volume per unit mass increases, since the
overall density of the grain is lowered. As a result, grains are more
efficient radiators of energy, and less dust mass in total is required
to reproduce an observed IR luminosity.

For a grain of the same mass, the grain is heated to a higher
temperature if it is solid than if it is porous. As a result of the
equilibrium temperature being lower for porous grains, more heating is
necessary to fit the same spectrum with a porous grain model than a
solid grain model. For a fixed plasma temperature, this can only be
achieved by increasing the post-shock density. In summary, increasing
the porosity of grains has the dual effects of raising the necessary
postshock density required to heat grains to an observed temperature,
while lowering the total amount of dust required to reproduce an
observed luminosity.

What remains is to choose a grain ``recipe'' to use to fit the
data. Since there are few constraints from the ISM at this time, we
have adopted here two values of $\cal P$, 25 and 50\%, to explore the
effects of grain porosity of fitting IR data. Our ``{\it compact}''
model, which we used to derive results given above, assumes separate
populations of solid silicate and graphite grains with appropriate
size distributions for the LMC (WD01). Our ``{\it porous}'' grain
models are based on those of \citet{clayton03}. For the 25\% porous
model, we assume that grains are composed of 25\% vacuum, 50\%
silicate, and 25\% amorphous carbon. The 50\% porous grains contain
50\% vacuum, 33.5\% silicate, and 16.5\% amorphous carbon, distributed
from 0.0025 to 1.5 $\mu$m. The narrow bandwidth of the IRS data does
not allow us to determine which model fits the data better, as we
obtained equally good fits from all models (albeit with a different
value of density), and report results from all. A caveat to using
these porous grain distributions is that both of them were optimized
for the Galaxy, and not the LMC. However, to the best of our
knowledge, a size distribution appropriate for porous grains in the
LMC has not yet been developed.

The results are as follows: For 0509, the best-fit $\cal P$ = 25\%
model had a post-shock density, $n_{H}$, of 1.1 cm$^{-3}$ ($\chi^{2}$
= 1.05), while the $\cal P$ = 50\% model required an $n_{H}$ of 2.2
cm$^{-3}$ ($\chi^{2}$ = 1.07). For 0519, the best-fit densities were
9.5 ($\chi^{2}$ = 1.08) and 17 ($\chi^{2}$ = 1.07) cm$^{-3}$ for $\cal
P$ = 25 and 50\%, respectively. In all cases, we assume LMC
abundances, such that $n_{e}$ = 1.2$n_{H}$. 

\section{X-ray Modeling of RGS Data}

Although dust models in the IR are a powerful diagnostic of the
post-shock gas, they are insensitive to pre-shock gas conditions,
which determine the total swept mass. Unlike many SNRs, both of these
objects have relatively well-defined ages and distances. The
post-shock densities are related to the pre-shock values by $r$, which
is 4 in the case of an unmodified, strong shock, but may be greater
than 4 in the case of cosmic-ray modification of the blast wave. For a
given $r$, we can calculate the pre-shock density, and thus the total
amount of swept up material. The product of post-shock electron
density and total gas mass overrun by the blast wave is proportional
to the emission measure (EM) of the X-ray emitting plasma. When
combined with the ionization timescale ($\tau_{i}=\int_0^t n_e dt$)
and electron temperature of the gas, we have all the information
necessary to generate a model X-ray spectrum for the shocked ambient
medium, which we can compare directly to observed X-ray spectra.

The difficulty in this approach lies in disentangling X-ray emission
from the shocked ambient medium from that arising from reverse-shocked
ejecta. For large remnants, this could be done spatially using CCD
spectra, but both 0509 and 0519 are $\sim 30''$ in diameter with the
ejecta not well separated from the shocked ISM, rendering such an
approach impossible. For 0509, previous studies
\citep{warren04,badenes08} have used Chandra CCD spectra to model
emission from the remnant, but these studies are subject to the
inherent difficulty in using poor spectral resolution CCD spectra to
disentangle ISM and ejecta emission. An alternative approach is to
take high-resolution spectra from grating spectrometers, such as RGS
on {\it XMM-Newton}. If lines can be identified as arising only from
shocked ambient medium, then model fits to these lines could be
directly compared with predicted line strengths from the model
described above. This can be most easily done with N lines, as
virtually no N is produced in Type Ia explosions. \citet{kosenko08}
fit the 24.8 \AA\ N Ly-$\alpha$ line in the spectrum of 0509, but we
find that this line is very weak and unconstraining to the fitting
procedure used here. Here we generate model O Ly$\alpha$ and K$\alpha$
lines visible in RGS spectra and compare them directly to observed
line fluxes. In Figure~\ref{rgsspectra}, we show RGS spectra from both
remnants.

\subsection{0509}

\begin{figure}
\includegraphics[width=12cm]{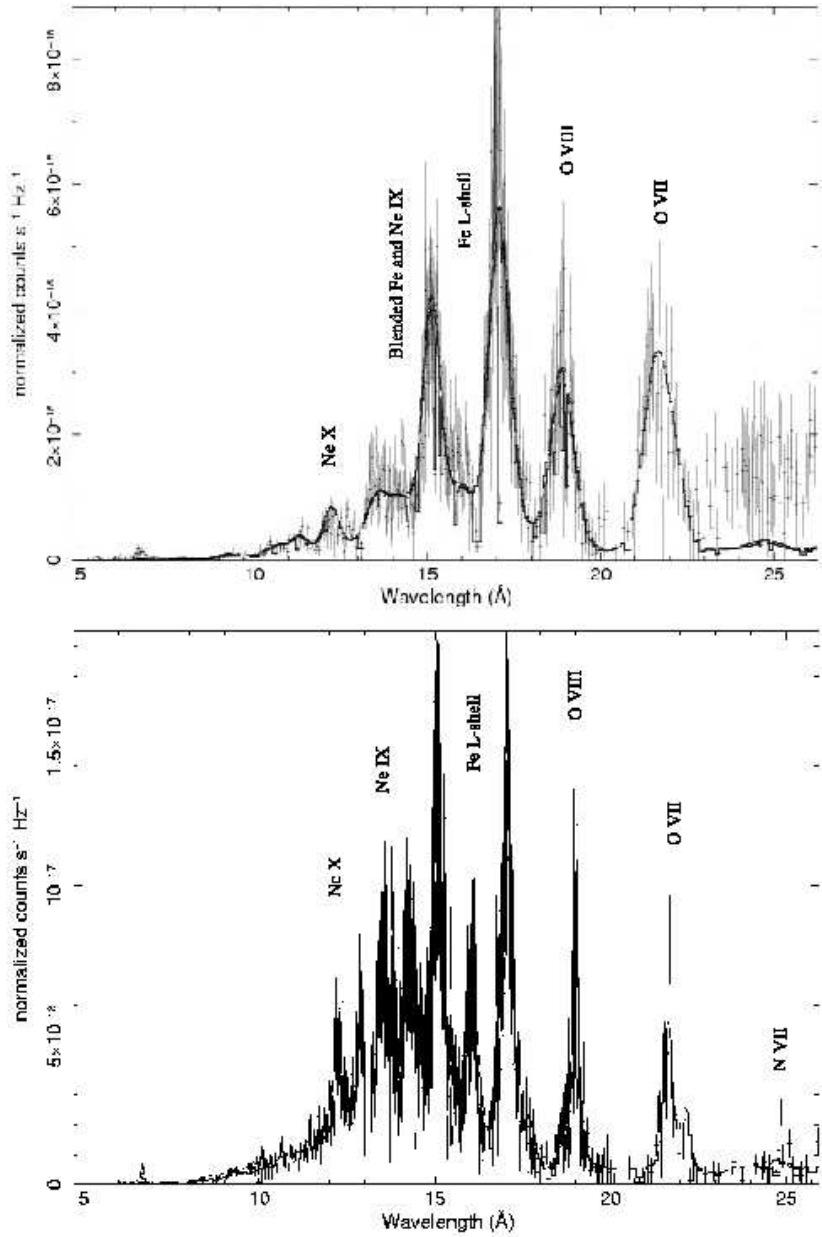}
\caption{Top: {\it XMM-Newton} RGS spectrum of B0509-67.5, from 5-27
\AA, with model overlaid as described in text. Lines of interest are
marked. High shock speeds and ejecta velocities lead to significant
blending of Ne lines with numerous Fe L-shell lines. Data beyond 23
\AA\ was excluded from model fitting, and is shown only for ease of
comparison with bottom panel. Bottom: Same, but for B0519-69.0. Slower
speeds allow separation of Ne and Fe L-shell lines in this remnant.
\label{rgsspectra}
}
\end{figure}

As seen in Figure~\ref{rgsspectra}, Fe L-shell lines dominate the
spectrum of 0509, but strong lines from both H and He-like O are
clearly visible in both cases at 18.97 and 21.7 \AA. In order to
measure the strength of observed O lines, we fit the data with a
two-component model. One model, a {\it vpshock} model containing only
iron, was fit to the data between 14-18 \AA, which are dominated by
iron lines. We then added a second {\it vpshock} model at fixed normal
LMC abundances ({\it wilms} abundance model in XSPEC \citep{wilms00},
with abundances of heavy elements set to 0.4; except that N = 0.1)
with ionization timescale and normalization free to fit the data
between 9-14 and 18-23 \AA, regions where the signal-to-noise was good
enough to constrain the fit. We use two absorption components, one
resulting from Galactic absorption with column density, $N_{H}$, equal
to 5 $\times$ 10$^{20}$ cm$^{-2}$ \citep{staveleysmith03}, and one
from the LMC (at fixed LMC abundances; same as above), equal to 2
$\times$ 10$^{20}$ cm$^{-2}$ \citep{warren04}. These models were used
only to determine the strength and width of oxygen lines.

We freeze the electron temperature for the second {\it vpshock} model
to 2 keV. This is roughly the number one gets from models of Coulomb
heating of electrons behind a shock of 7000 km s$^{-1}$ with a
preshock density of 0.25 cm$^{-3}$. Although we do not know the
preshock density {\it a priori}, we use this as a representative value
and note that the dependence of the emission measure of the gas on the
electron temperature is small. We introduce an additional line
smoothing parameter to both models to allow for the fact that lines
are broadened by Doppler motions. For this model, we find that oxygen
line fluxes measured from RGS spectra in 0509 are as follows: O
Ly$\alpha$ = 4.41 $\times 10^{-13}$ ergs cm$^{-2}$ s$^{-1}$ and O
K$\alpha$ = 5.05 $\times 10^{-13}$ ergs cm$^{-2}$ s$^{-1}$. The oxygen
line widths obtained from fitting are $v_{FWHM}$ = 12350 (11220,
13630) km s$^{-1}$. \citet{kosenko08} measured the widths of Fe lines
in the RGS spectrum of this object, obtaining line widths of 11500
(10500, 12500) km s$^{-1}$.

\subsection{0519}
\label{0519xray}

Our technique for fitting RGS data from 0519 was identical, except
that we extended the fitting range out to 27 \AA\ to account for the
fact that we had better signal-to-noise at long wavelengths in this
case. We again used two absorption components; one for the galaxy with
$N_{H}$ = 6 $\times$ 10$^{20}$ cm$^{-2}$ \citep{staveleysmith03} and
one for the LMC with $N_{H}$ = 1.6 $\times$ 10$^{21}$ cm$^{-2}$
\citep{hughes95,kosenko10}. Measured line strengths for this remnant
are 5.81 $\times 10^{-13}$ ergs cm$^{-2}$ s$^{-1}$ and 3.91 $\times
10^{-13}$ ergs cm$^{-2}$ s$^{-1}$ for O Ly$\alpha$ and O K$\alpha$,
respectively. We obtain a line width for oxygen lines of $v_{FWHM}$ =
3470 (3170, 4000) km s$^{-1}$. Using a three-component NEI model,
\citet{kosenko10} measure Fe line widths in this object to be 4400
(4280, 4520) km s$^{-1}$. The also report that neither Ne nor Mg is
required to fit the {\it XMM} spectra. We find that normal abundances
of Ne and Mg from the shocked ISM are allowed by the data, and it is
not necessary to exclude them from the fitting procedure.

Oxygen line widths in 0519 are comparable to the assumed shock
velocity, while those in 0509 are significantly higher. In
Section~\ref{hydro}, we discuss hydrodynamic modeling of 0519, and
find that given the age of 0519, a lower shock speed of 2450 km
s$^{-1}$ is preferred to match the observed radius of the forward
shock. Line widths from a shock of this speed would fall within errors
of the observed line widths above. 

\section{Discussion}
\label{disc}

IR spectral fitting gives us a postshock density, which we can then
divide by $r$ of the shock to get the preshock density. For a
standard, ``strong'' shock, $r$ is $\rho_{1}/\rho_{0} = 4$, but
modification of shocks by cosmic ray acceleration will increase
this. An additional complication for the cases of these two objects is
that they are both only 30$''$ in diameter, which is quite small for
the telescope resolution at {\it Spitzer} wavelengths, meaning that
the spectra shown in Figures 2-4 are the average spectra over the
entire postshock region. To account for this, we computed hydrodynamic
models (see Section~\ref{hydro}) and calculated the mean, EM-weighted
$r$ over the post-shock region from the leading edge of the blast wave
to the contact discontinuity. For the case of a standard shock, this
mean $r$ is 2.65. This is lower than the immediate postshock value of
4 due to the sphericity of the blast wave. The postshock density drops
behind a spherical shock.

\begin{figure}
\includegraphics[width=14.5cm]{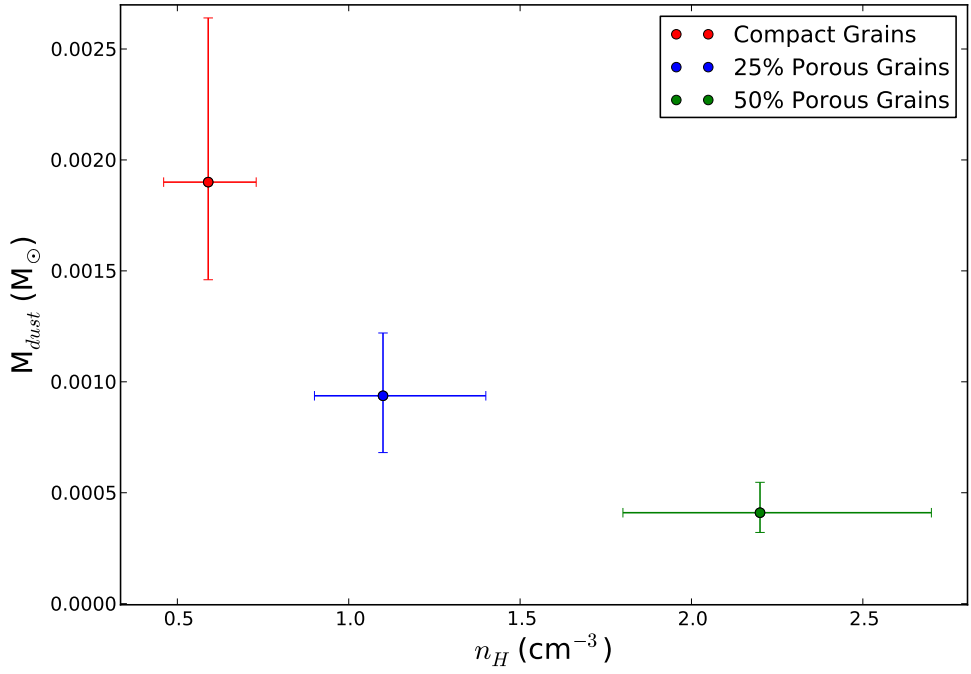}
\caption{Best fit values for 0509 of post-shock density and dust
mass, with errors, for various grain models. Errors are 90\%
confidence limits, as described in the text.
\label{0509_2dplot}
}
\end{figure}

\begin{figure}
\includegraphics[width=14.5cm]{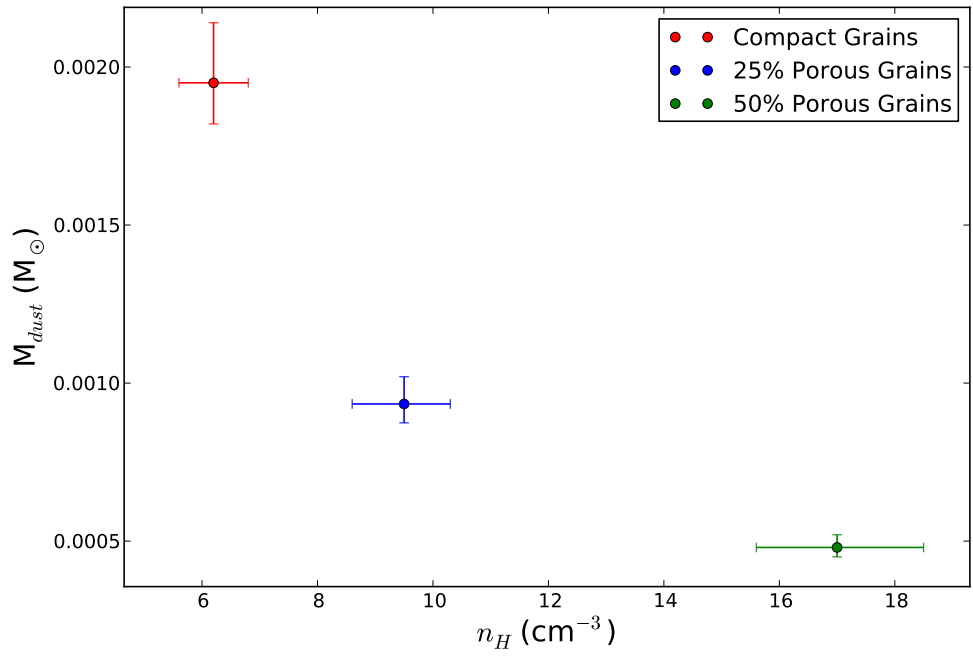}
\caption{Same as Figure~\ref{0509_2dplot}, but for 0519
\label{0519_2dplot}
}
\end{figure}

Taking the example of 0509, dust model fits to IR spectra give a
postshock density of $n_{H} = 0.59$ cm$^{-3}$. Dividing this by 2.65,
we get an ambient pre-shock density, $n_{0}$, of 0.22 cm$^{-3}$. The
known distance to the LMC of 50 kpc, combined with the measured
angular size of the remnants, gives a shocked volume of 5.7 $\times
10^{57}$ cm$^{3}$. Multiplying the pre-shock density by the volume
gives the value reported in table 2 for the swept-gas mass, 1.47
$M_\odot$. Dividing the dust mass inferred from fits to IR data, 1.9
$\times 10^{-3}$ $M_\odot$, by this number, we arrive at an ambient
dust-to-gas mass ratio of 1.3 $\times 10^{-3}$, the number we report
in Table 1. This dust-to-gas ratio is not folded into the models at
any point, nor is it ever assumed. We repeat these steps for various
values of $r$, determining different swept gas masses and dust-to-gas
mass ratios. However, note that the post-shock density and total dust
mass remain constant for each case, since those were determined from
fits to IR data. We include effects of sputtering of dust grains by
ions in deriving the dust masses above. In Figures~\ref{0509_2dplot}
and~\ref{0519_2dplot}, we show the best fit values of post-shock
density and dust mass for various grain models.

X-rays in 0509 and 0519 are coming mostly from reverse-shocked
ejecta. The shocked ambient medium may contribute substantially,
particularly to continua and O lines, though it is difficult to
evaluate this contribution based solely on X-ray data. However,
because fits to IR data provide us with the post-shock density, and
the ages and sizes of these remnants are well-known, we can use X-ray
emission codes to predict line strengths coming from the shocked
ambient medium, assuming typical LMC abundances.

In Tables 2 \& 3, we show predicted oxygen line strengths as a
function of $r$ and grain porosity for both remnants. These line
strengths were calculated in {\it XSPEC} by taking the plane-shock
models as described in the section above and varying the ionization
timescale and emission measure appropriately for each case. Real ages
of 400 and 600 years for 0509 and 0519, respectively, were divided by
3 in calculating $\tau_{i}$ as a correction factor for applying a
plane-shock model to a spherical blast wave \citep{borkowski01}. EM
was calculated as $n_{e} n_{0} V$, where $n_{e}$ is post-shock
electron density, $n_{0}$ is pre-shock density ($\equiv$
$\frac{n_{H}}{r}$, where V is total volume enclosed by the blast wave
(approx. $5.7 \times 10^{57}$ cm$^{3}$ for both remnants). We chose
values for $r$ of 4, 8, and 12, which correspond to mean EM-weighted
$r$ values of 2.65, 4.8, and 6.8.

In the Tables, we quote our predicted O line strengths from the blast
wave as a fraction of the observed strengths from RGS spectra. Values
$<$ 1 imply a contribution to O lines from ejecta, while values $>$ 1
are inconsistent with observations and are ruled out. For 0519, this
rules out all models of porous grains. However, this requirement also
rules out compact grain models that are unmodified from ``standard''
shock jump conditions.  Only those shocks with significantly higher
$r$ are possible based on X-ray line strengths observed.

For 0509, only one case is ruled out, that of highly porous ($\cal P$
= 50\%) grains heated behind a shock with standard jump conditions. If
grains are compact, then virtually all O line emission seen in RGS
spectra would have to come from ejecta. Recent near-IR observations of
Type Ia SNe \citep{marion09} suggest that the entire progenitor is
burned in the explosion and that O and Ne are byproducts of carbon
burning, meaning that strong X-ray lines from oxygen in the ejecta may
be unlikely. The favored explosion model for 0509 from
\citet{badenes08} contains 0.04 $M_\odot$ of unburned oxygen, the
least of any of the explosion models considered for this
remnant. Nevertheless, even this relatively small amount of ejecta
oxygen could be enough to produce observed O lines. Model line ratios
require at least some contribution from reverse-shocked ejecta to
account for observed O Ly$\alpha$ strength.

\citet{helder10} find that at least some of the energy of the shock in
0509 must be deposited into cosmic rays. Of course, if cosmic-ray
pressure is indeed non-negligible, then the standard case no longer
applies, and cosmic-ray modified shocks would be required. Our assumed
proton temperature of 70 keV already implies modification from the
standard shock jump conditions, which would predict $kT_{p}$ $\sim
115$ keV. However, the dependency of the dust heating models on this
number is small (see Section~\ref{iontemp}). The relationship between
cosmic-ray pressure and shock compression ratio is model dependent
\citep{vink10} and beyond the scope of this paper, but models in Table
1 with $r > 4$ would be preferred.

The data do not allow us to firmly distinguish between compact and
porous grain fits, as both can be consistent with the data under the
right circumstances. Instead, we summarize implications for each
model. If dust grains in the ISM are compact, the oxygen lines visible
in X-ray spectra of 0509 are dominated by contributions from the
ejecta, for all values of $r$. The inferred preshock density is
inversely proportional to $r$ (see Table~\ref{resultstable0509}). A
``standard'' $r$ value of 4 yields a preshock density of $n_{0}$ =
0.22 cm$^{-3}$, significantly higher than either the value of 0.05
cm$^{-3}$ quoted by \citet{warren04} or that of 0.02 cm$^{-3}$ from
\citet{tuohy82}, but below the 0.43 cm$^{-3}$ of \citet{badenes08}. In
0519, it is possible for nearly all of O lines to be coming from
shocked ambient medium; however, the minimum $r$ value required to
match the O K$\alpha$ line strength observed is 12. For this value of
$r$, a small ejecta contribution to the O Ly$\alpha$ line is still
necessary for line ratios to match.

If grains are porous, a lower ejecta contribution to oxygen lines
observed in 0509 would be permitted, even for values of $r$ that are
closer to the standard shock value. However, in 0519, much higher $r$
values would be required to bring model oxygen line fluxes down to
observed levels, even with no contribution from the ejecta. We require
significantly less oxygen in the ejecta than \citet{kosenko10}, who
report an O ejecta mass of 0.2-0.3 $M_\odot$. Minimum values for $r$
would be 20 and 34 for $\cal P$ = 25\% and 50\%, respectively.

\subsection{Radio Limits on Compression Ratio}
\label{radiolimits}

We can set an upper limit to $r$ using the observed diffuse
synchrotron brightness of the LMC near 0509 and 0519, and requiring
that the brightening by compression of the ambient synchrotron
emissivity not exceed the observed radio brightnesses of 0509 and
0519.  Compression will boost the energy of electrons and also
increase the mean magnetic field strength in the emissivity $j_\nu$.
If the electron energy distribution is $N(E) = KE^{-s}$ between
energies $E_l$ and $E_h$, then the emissivity is given by
\begin{equation}
j_\nu = c_j(\alpha) K B^{1 + \alpha} \nu^{-\alpha}
\end{equation}
where $\alpha \equiv (s - 1)/2$ and $c_j(0.5) = 4.85 \times 10^{-14}$
cgs.  (In general, $c_j \equiv c_5(\alpha) (2c_1)^\alpha$ in the
notation of Pacholczyk [1970], with $c_1 \equiv 1.82 \times 10^{18}$
and $c_5(0.5) = 1.37 \times 10^{-23}$.)
Taking $\alpha = 0.5$ $(s = 2)$ for simplicity, we find that the
electron energy density $u_e$ obeys 
\begin{equation}
u_e \equiv \int_{E_l}^{E_h} KE^{-s} E dE = K\ln\left( {E_h \over E_l} \right)
\propto p_e \propto \rho_e^{4/3} \propto r^{4/3}.
\end{equation}

In the absence of turbulent effects and magnetic-field
amplification,the tangential component of $B$ will be increased by $r$
while the radial component will remain the same, raising the magnitude
of $B$ by a factor $r_B < r$.  For a spherical remnant encountering a
uniform magnetic field, it is straightforward to show that the mean
amplification factor $\langle{r_B \rangle}$ is $0.79r$ for $r \ge 6$.
Thus, after compression, the synchrotron emissivity interior to the
remnant is given by
\begin{equation}
{j_\nu ({\rm in}) \over j_\nu({\rm out})} = r^{4/3} (0.79 r)^{1.5} 
= 0.70 r^{17/6}.
\end{equation}

For a mean diffuse surface brightness upstream of $\Sigma_0$ W
m$^{-2}$ sr$^{-1}$, the mean emissivity in the LMC just depends on the
line-of-sight depth of synchrotron-emitting material in the LMC.
Calling that quantity $l$, we have $\langle{j_\nu({\rm up})\rangle} =
\Sigma_0/l$.  Compressing this as above predicts the post-shock
surface brightness for a SNR of radius $R$, $I_{\rm SNR}$, to be about
\begin{eqnarray}
I_{\rm SNR} = j_\nu({\rm in}) R = 0.70 r^{17/6} j_\nu({\rm out}) R
= 0.70 r^{17/6} \left( \Sigma_0 \over l\right) R\\
\Rightarrow r_{\rm max} = 1.13 \left( I_{\rm SNR} \over \Sigma_0\right)^{6/17}
\left(l \over R\right)^{6/17}.\\
\end{eqnarray}

The diffuse radio brightness of the LMC has been mapped with ATCA and
Parkes (Hughes et al.~2007).  From the image at 20 cm (resolution
$40''$), we estimate surface brightnesses near 0509 and 0519 of about
1 and 2 mJy/beam respectively, or about $1 \times 10^{-21}$ and 
$2 \times 10^{-21}$ W m$^{-2}$ sr$^{-1}$ at 1 GHz, respectively
(using a mean spectral index of the diffuse emission of 0.3; Hughes
et al.~2007).  We take $l \sim 1$ kpc.  

For 0519, a high-resolution image is available \citep{dickel95}, which
shows a shell peak brightness of about 6 mJy/beam at 20 cm, giving
(for $\alpha = 0.5$ and a FWHM of $4.8''$) $I_{\rm SNR} \sim 1 \times
10^{-19}$ W m$^{-2}$ sr$^{-1}$, and $r_{\rm max} \sim 26$.  There is
no high-resolution image of 0509 available; using the mean remnant
surface brightness ($\sim 5 \times 10^{-20}$ W m$^{-2}$ sr$^{-1}$ from
data in Mills et al.~1984) slightly overestimates $r_{\rm max}$, but
gives a value of 25.  Larger values of $r$ would give higher radio
brightnesses than we observe.

\subsection{Effect of Decreased Ion Temperature on IR Fitting}
\label{iontemp}

If shocks are indeed modified by cosmic-ray accleration, both $r$ and
the plasma temperature behind the shock would be altered from the
standard strong shock jump conditions
\citep{jones91}. \citet{helder10} report evidence in 0509 for
decreased proton temperatures due to cosmic-ray modification of the
shock. In this work, we find that higher than normal values of $r$ are
also necessary to avoid overpredicting oxygen lines in the X-ray
spectra of 0519; such a modification would likewise lower proton
temperatures from the value assumed. Dust model fits to IR data are
dependent only on these post-shock conditions, and are insensitive to
the degree of modification at the shock. A detailed examination of
this relationship is beyond the scope of this paper, but we can make
some general comments about the effect that lowering the proton
temperature behind the shock would have on fitting the data. In the
case of most SNRs, heating of grains by protons can be neglected, as
heating is dominated by collisions with electrons. However, for
remnants with very fast shocks, proton heating becomes non-negligible
\citep{williams10}.

In Table~\ref{resultstable0509}, we use a proton temperature of 70 keV
for modeling 0509. We take the case of 25\% porous grains as an
example (where $n_{H}$ was fit to be 1.1 cm$^{-3}$). Raising the
proton temperature by a factor of $\sim 50\%$ to 115 keV, the
temperature expected from a 7000 km s$^{-1}$ shock with no
electron-ion equilibration lowers the fitted density to 1.0 cm$^{-3}$,
an effect of only about 10\%. Lowering by a factor of $\sim 10$ to
$T_{p} = 7.5$ keV raises $n_{H}$ to 1.3 cm$^{-3}$. Finally, setting
the proton temperature to $T_{p}$ = $T_{e}$ = 2 keV (a lower limit;
although unphysical in this object given that \citet{ghavamian07}
report detection of broad Ly$\beta$ UV line with width 3710 km
s$^{-1}$) requires $n_{H}$ = 2.0 cm$^{-3}$. These effects are
non-negligible, but are not enough to affect the overall conclusions
stated above.

\subsection{Hydrodynamic Modeling of 0519-69.0}
\label{hydro}

We can derive a conservative lower limit on the pre-shock density in
0519 using one of two methods. The H$\alpha$ fluxes from
\citet{tuohy82} give a firm lower limit on $n_{0}$ of 0.2 cm$^{-3}$ at
our assumed shock speed into a fully neutral medium, and assuming 0.2
H$\alpha$ photons per atom entering the shock. The radio upper limit
to $r$ (see Section~\ref{radiolimits}) of 25, which translates to a
mean $r$ of $\sim 12.5$, combined with the minimum post-shock density
of 6.2 cm$^{-3}$ yields a lower limit on $n_{0}$ of 0.5 cm$^{-3}$. An
upper limit on the pre-shock density in 0519 is given in Table 3 as
0.91 cm$^{-3}$, which is a factor of $\sim 2.5$ lower than the
pre-shock density of $n_{0}$ = 2.4 cm$^{-3}$ quoted by
\citet{kosenko10}, who derive their ambient density by fitting a
six-component NEI model to the X-ray spectra from {\it Chandra} and
{\it XMM-Newton}. In their model, most of the oxygen comes from the
ejecta, which requires a very high ionization timescale of $8 \times
10^{11}$ cm$^{-3}$ s (a factor of $\sim 20$ higher than our value
reported in Table 3) and thus a very high EM ($2.36 \times 10^{59}$
cm$^{3}$) for the shocked CSM component, implying a higher pre-shock
density.

\begin{figure}
\includegraphics[width=14cm]{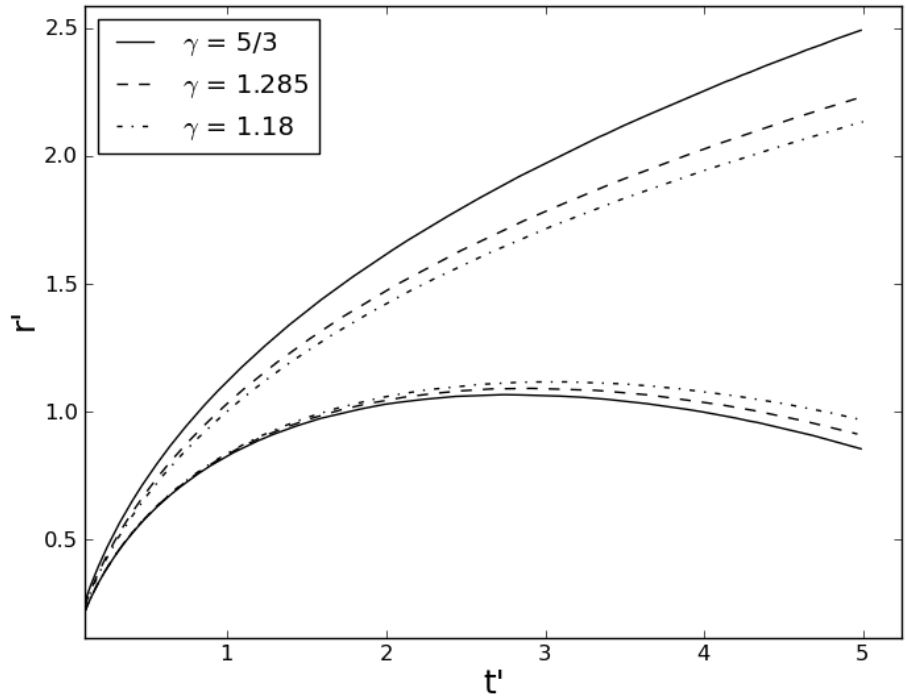}
\caption{Evolution of the forward and reverse shocks with time for
several values of the adiabatic index, $\gamma$, of the shocked
medium. Top curves correspond to the forward shock; bottom curves
correspond to the reverse shock. $t'$ and $r'$ are normalized to
values given in \citet{dwarkadas98}. Solid line here corresponds to
the solid line in Figure 2a of their paper. Note that decreasing
$\gamma$ has the effect of shrinking the post-shock volume. Effective
$\gamma$ values of 5/3, 1.285, and 1.18 correspond to effective $r$
values of 2.65, 4.8, and 6.8, respectively.
\label{fsrs}
}
\end{figure}

There is roughly a factor of 2-3 between our lower and upper limits to
the pre-shock density. To explore both of these limits, we use the
Virginia Hydrodynamics (VH-1) time-dependent hydrodynamic numerical
code in one dimension. VH-1 is a hydrodynamics code which solves the
Eulerian equations of fluid mechanics using the Piecewise Parabolic
Method, described in \citet{colella84}. We assume spherical symmetry
on a numerical grid with 512 zones. We modify $r$ for the forward
shock by varying the adiabatic index, $\gamma$ (= $\frac{5}{3}$ for a
normal shock), of the shocked ambient medium
\citep{blondin01}. Softening the equation of state effectively makes
the shocked plasma more compressible, yielding $r$ $>$ 4. We use the
exponential ejecta density profiles of \citet{dwarkadas98} and assume
an explosion energy of $10^{51}$ ergs with an ejected mass of 1.4
$M_\odot$. In Figure~\ref{fsrs}, we plot the evolution of the forward
and reverse shocks with time for varying values of $\gamma$,
corresponding to immediate postshock $r$ values of 4, 8, and 12, using
identical scalings to that of \citet{dwarkadas98} (compare with Figure
2a of their paper).

With the mean post-shock density fixed at 6.2 cm$^{-3}$ (the gas
density required to fit the spectrum assuming compact grains), we
examined $r$ values of 12 and 25, corresponding to $\gamma$ values of
1.18 and 1.085, mean $r$ in the post-shock region of 6.8 and 12.5, and
pre-shock densities of 0.91 and 0.50, respectively.

The lower density case, with $r = 25$, reaches a radius of 3.6 pc in
550 years with a shock velocity at that time of 2950 km
s$^{-1}$. While this age is roughly consistent with the known age of
this object from light echoes, the shock velocity is too high to be
consistent with the measured width of oxygen lines. This shock speed
would imply a width, $v_{FWHM}$, of 5660 km s$^{-1}$. The measured
value was $\sim 3500$ km s$^{-1}$. The higher density case requires
630 years to reach the observed radius, with a current shock speed of
2450 km s$^{-1}$. The line width implied from this shock speed is 4490
km s$^{-1}$. While this is still higher than our observed line width,
it is much closer to being within the quoted errors, and is clearly
favored over the lower density case.

Errors in determining the post-shock density from IR fits are about
20\%, but this has almost no effect on the inferred pre-shock density
allowed by not overpredicting oxygen lines in RGS spectra. However,
this number is sensitive to the soft X-ray absorbing column density
used in the models. If the LMC absorption component, detailed in
Section~\ref{0519xray}, is increased, a higher pre-shock density will
be allowed before oxygen lines are overproduced. To be within the
errors of the line widths measured for oxygen lines, the LMC $N_{H}$
value would only need to be modestly increased from 1.6 to 1.7 $\times
10^{21}$ cm$^{-2}$. \citet{kosenko10} report an $N_{H}$ of 2.6 $\times
10^{21}$ cm$^{-2}$, which would easily accommodate a higher
density. Uncertainties in the exact distance to the remnant are also
non-negligible here. We assume a distance to the LMC of 50 kpc, but
recent measurements of the distance to the LMC range from 47.8 kpc
\citep{grocholski07} to 51.1 kpc \citep{koerwer09}. The lower of these
two would increase the observed radius of the remnant from 3.6 to 3.77
pc, which would bring the observed line width within errors of the
shock velocity from hydro modeling. The unknown position of 0519
within the depth of the LMC may also be important.

\subsection{Dust-to-Gas Mass Ratio}
\label{dustmass}

Of particular note in all results presented in Tables 2 and 3 is that
given the amount of gas that these remnants have swept up at this
stage of their evolution, the amount of dust observed by {\it Spitzer}
is lower than what is expected, even accounting for the effects of
sputtering (although there are a few values in Table 2 that are
acceptable; namely, those with a high $r$). The amount of dust
sputtered is $\sim 10$\% for compact grains in 0509, and 34\% for
compact grains in 0519. UV studies of grain absorption in the ISM
\citep{weingartner01} yield a dust-to-gas mass ratio in the LMC of 2.5
$\times 10^{-3}$. Recent wide-field far-IR observations of the LMC
give a value of $\sim 3.5 \times 10^{-3}$ for this ratio
\citep{meixner10}. This shortfall is consistent with our previous
studies of both LMC and galactic SNRs
\citep{borkowski06,williams06,blair07}. Similar results have been
found for SN 1987A \citep{bouchet06} in the LMC, although the shocks
there probe circumstellar material, and have not yet reached the
ambient ISM (but see Dwek et al. 2008,2010), and for G292.0+1.8
\citep{lee09} and Puppis A \citep{arendt10} in the Galaxy.

Of all parameters in the model, dust mass is the {\it least}
model-dependent. As a check of this, we examine the case of
0519. Assuming compact grains, we derive a total radiating dust mass
of 1.5 $\times 10^{-3} M_\odot$ based on model fits to IRS data. We
then made fits to the data using a greatly simplified model, that of a
single grain at a single temperature, ignoring heating, sputtering,
and grain-size distribution. This model yielded a total dust mass of
about 1.3 $\times 10^{-3} M_\odot$. As a further check, we used an
approximate analytic expression from \citet{dwek87} for the mass of
dust, dependent only on the temperature of the grain and the observed
IR luminosity, obtaining a value of 1.7 $\times 10^{-3}
M_\odot$. These three different approaches of decreasing complexity
only differ by at most a factor of 30\%. Taking our {\it minimum}
value of swept-up gas mass (5.3 $M_\odot$), we derive dust-to-gas mass
ratios of 2.4-3.2 $\times 10^{-4}$, an order of magnitude lower than
the expected value. Dust-to-gas mass ratios in 0509 are closer to
expected values, but are, in most cases in
Table~\ref{resultstable0509}, still too low. It may be that SNRs that
are in regions of low background sufficient to allow a high enough
signal-to-noise ratio for a detailed study in the mid- and far-IR have
such a low background because they are in atypical regions of the ISM,
creating an observational bias. In addition, dust-to-gas mass ratio
determinations from SNRs are {\it spot} measurements, and are not
averaged over long lines of sight or large fields of view.

Porous dust grains require a higher density relative to compact grains
to reproduce the same spectrum, because they are more efficient
radiators of energy than compact grains. For this reason, the inferred
dust mass from the porous grain models, seen in Table 2, is lower by a
factor of $\sim 2-4$. However, for a given $r$, the inferred gas mass
is higher for porous grains, so that the dust-to-gas mass ratios for
porous models are lower by a factor of $\sim 4$ for $\cal P$ = 25\%
and $\sim 18$ for $\cal P$ = 50\%, when compared to the compact grain
case. It should be noted though, that typical dust-to-gas mass ratios
calculated for the ISM in general (such as the 0.25\% given above from
WD01) inherently assume compact grains.

One possibility that has been suggested for SNRs of both Type Ia and
CC origin \citep{gomez09,dunne03} is that large amounts of dust are
indeed present, but are too cold (T$_{d}$ $\sim 16$ K) to be
detectable at {\it Spitzer} wavelengths and are only visible in the
sub-millimeter regime. If this is true, then the total dust-to-gas
mass ratio for the ISM could be quite different from what we report
here, which is based only on ``warm'' dust. This claim has been
disputed for Kepler's SNR \citep{blair07} based on a detection at 70
$\mu$m and an upper limit at 160 $\mu$m, but we lack detection at
either wavelength here, and upper limits are unconstraining. {\it
Herschel} observations would provide the key wavelength coverage
between the mid-IR and sub-mm necessary to settle this issue. A recent
{\it Herschel} observation of Cas A at wavelengths up to 500 $\mu$m
revealed only 0.075 $M_\odot$ of dust there, much of which may be
newly-formed ejecta dust \citep{barlow10}.

How cold would dust have to be to escape detection at 70 $\mu$m,
assuming a standard dust-to-gas mass ratio for the LMC? Obviously, the
answer depends on the choice of models from Tables 2 \& 3. As an
example, we consider the $\cal P$ = 0 model for 0519 with the highest
value for $r$ (the only model for this remnant that seems to not be
ruled out by X-ray lines). The swept gas mass for this case is 6.1
$M_\odot$, and assuming a dust-to-gas ratio of 0.35\% gives a dust
mass of 2.1 $\times 10^{-2}$ $M_\odot$. In order to avoid violating
the 70 $\mu$m upper limit of 121 mJy \citep{borkowski06}, post-shock
dust in this remnant would need to be $\lapprox 44$ K. If we choose
the standard $r$ case, the implied dust mass would be 5.5 $\times
10^{-2}$ $M_\odot$, and the required post-shock dust temperature would
be $\lapprox 36$ K. Both of these numbers are significantly lower than
those fit to IRS data or those predicted by grain heating models. Such
low temperatures could only be acheived by having the dust reside in
very dense clumps. Shocks driven into such clumps should quickly
become radiative, but no [O III] or [S II] emission is seen from
either remnant.

\section{Conclusions}

We present mid-IR spectral observations of two young SNRs in the LMC,
0509-67.5 and 0519-69.0, as well as analysis of archival
high-resolution X-ray spectra. By fitting dust heating and sputtering
models to IR spectra, we can determine post-shock gas density. We
derive post-shock densities of 0.59 and 6.2 cm$^{-3}$ in 0509 and 0519,
respectively, using a compact grain model. Porous grains require a
density higher by a factor of a few. By assuming a value for the shock
compression ratio, $r$, we can infer the pre-shock density, swept gas
mass, strength of oxygen Ly$\alpha$ and K$\alpha$ X-ray lines, and
dust-to-gas mass ratio for the ambient ISM. Derived values for the
pre-shock density of the ISM vary greatly depending on the model used
and compression ratio assumed, but for the sake of comparison, values
listed in Tables 2 \& 3 are comparable to densities inferred from
various line-of-sight HI column densities through the LMC. These
column densities vary from 10$^{20}$ cm$^{-2}$ to a few $\times
10^{21}$ cm$^{-2}$, which, assuming an LMC depth of 1 kpc, yields
densities of $\sim 0.05-1.5$. For standard strong shocks, $r$ is 4,
but modification by cosmic rays would increase this by an unknown
amount. We report values of the quantities above for several plausible
values of $r$. In principle, this method could be used in reverse to
determine $r$. This would require X-ray emission from shocked ambient
medium that is well-separated from ejecta emission; thus, X-ray lines
could be modeled as solely arising from material at cosmic
abundances. We believe this method can be used for older, larger
remnants in the LMC, such as DEM L71.

Based on our analysis, a significant fraction of the O line emission
seen in the X-ray spectrum of 0509 arises from the ejecta. If grains
are highly porous, then the ejecta contribution is less, but a
contribution is still required because the line flux ratios observed
still do not match the model ratios. Both standard shocks and
cosmic-ray modified shocks with higher $r$ values can provide
acceptable fits for 0509. The data for 0509 are inconclusive in
favoring either compact or porous grains. Compact grains require
nearly all of the oxygen seen in X-ray spectra to be coming from
shocked ejecta, while porous grains would significantly lower the
ejecta contribution, implying a higher contribution from the
forward-shocked material.

In 0519, there is significant evidence for higher than standard
compression, and little contribution from ejecta to O X-ray lines. A
shock speed of $<$ 2300 km s$^{-1}$ is favored for this object from
hydrodynamical simulations, in order to reproduce the observed size,
age, and X-ray line widths. Compact grains are favored for this
remnant. The standard value of $r=4$ is possible only if absorption to
the remnant is significantly higher than has been reported. The column
density from the LMC would need to be at least $4 \times 10^{21}$
cm$^{-2}$ to bring oxygen lines in the model down to below observed
values. \citet{ghavamian07} derive an upper limit for 0519 of $E(B-V)
\le 0.11$, implying an upper limit to the HI column density of $\sim
1.6$ $\times 10^{21}$ cm$^{-2}$ \citep{cox06}.

We derive dust-to-gas mass ratios that are lower by a factor of
several than what is generally expected in the ISM. Since the general
properties of dust in the ISM come from optical/UV absorption line
studies averaged over long lines of sight, it is possible that there
exist large local variations on smaller scales. The ratios presented
here probe parsec scales. Another potential explanation is that the
sputtering rate for dust grains is significantly understated in the
literature. However, since the deficit of dust is about an order of
magnitude, sputtering rates would also have to be increased by this
amount to account for the discrepancy. Also, because sputtering rates
typically assume a compact grain, further work in this field will be
needed to determine if these rates are appropriate for porous
grains. Extremely high values for $r$ ($>$ 50) could bring dust-to-gas
mass ratios up to more typical values, but such values are ruled out
by radio observations. A possible observational bias could exist, in
that SNRs that are easy to study in the IR (i.e., well separated from
sources of IR confusion) could be found preferentially in low-dust
regions of the ISM.

Studies such as this will benefit greatly from the increases in both
spatial resolution and sensitivity of future generations of
telescopes, such as the {\it James Webb Space Telescope}. Being able
to spatially separate the dust spectra right behind the shock from
that further inside the shell will be crucial to reducing some of the
uncertainties listed above. Longer wavelength observations with the
{\it Herschel Space Observatory} would be useful for detecting or
placing upper limits on any cold dust that may be present in the
remnant.

\acknowledgments

We thank John Blondin for providing the VH-1 hydrodynamics code and
for useful discussions on interpretation, and the anonymous referee
for many useful comments which improved the paper. We acknowledge
support from {\it Spitzer} Guest Observer Grant RSA 1328682.

{\it Facilities:} \facility{XMM}, \facility{Spitzer}

\newpage
\clearpage

\begin{deluxetable}{lccccc}
\tablecolumns{6}
\tablewidth{0pc}
\tabletypesize{\footnotesize}
\tablecaption{Model Input Parameters}
\tablehead{
\colhead{Object} & $V_{s}$ (km s$^{-1}$) & $T_e$ (keV) & $T_p$ (keV) & Age (yrs.) & Ref.}

\startdata

B0509-67.5 & 6000 & 2.0 & 70 & 400 & 1, 2\\
B0519-69.0 & 3000 & 1.5 & 21 & 600 & 1, 2\\

\enddata

\tablecomments{References: (1) Ghavamian et al 2007, (2) Rest et al. 2005}
\label{inputtable}
\end{deluxetable}

\newpage
\clearpage

\begin{deluxetable}{lccccccccc}
\rotate
\vspace{-0.3truein}
\tablecolumns{10}
\tablewidth{0pc}
\tabletypesize{\footnotesize}
\tablecaption{Predicted ISM Oxygen Line Strengths for SNR B0509-67.5}
\tablehead{
\colhead{$\cal P$} & $n_{H}$\tablenotemark{a} & $n_{0}$\tablenotemark{a} & $n_{e}t$\tablenotemark{b} & EM\tablenotemark{c} & O Ly$\alpha$ & O K$\alpha$ & M$_{G}$\tablenotemark{d} & M$_{D}$\tablenotemark{e} & M$_{D}$/M$_{G}$ }

\startdata

0 & 0.59$^{0.73}_{0.46}$ & 0.22$^{0.28}_{0.17}$ & 2.97$^{3.67}_{2.31}$ $\times 10^{9}$ & 0.089$^{0.14}_{0.053}$ & 0.026$^{0.044}_{0.015}$ & 0.1$^{0.17}_{0.05}$ & 1.47$^{1.82}_{1.15}$ & 1.90$^{2.64}_{1.46}$ $\times 10^{-3}$ & 1.3$^{2.3}_{0.8}$ $\times 10^{-3}$ \\

0 & 0.59$^{0.73}_{0.46}$ & 0.12$^{0.15}_{0.1}$ & 2.97$^{3.67}_{2.31}$ $\times 10^{9}$ & 0.048$^{0.075}_{0.031}$ & 0.017$^{0.026}_{0.012}$ & 0.054$^{0.089}_{0.032}$ & 0.8$^{0.99}_{0.62}$ & 1.90$^{2.64}_{1.46}$ $\times 10^{-3}$ & 2.4$^{4.3}_{1.5}$ $\times 10^{-3}$ \\

0 & 0.59$^{0.73}_{0.46}$ & 0.09$^{0.11}_{0.068}$ & 2.97$^{3.67}_{2.31}$ $\times 10^{9}$ & 0.036$^{0.055}_{0.021}$ & 0.014$^{0.021}_{0.01}$ & 0.041$^{0.066}_{0.022}$ & 0.6$^{0.74}_{0.47}$ & 1.90$^{2.64}_{1.46}$ $\times 10^{-3}$ & 3.2$^{5.6}_{1.9}$ $\times 10^{-3}$ \\
\\
\hline
\\
25\% & 1.1$^{1.4}_{0.9}$ & 0.42$^{0.53}_{0.34}$ & 5.53$^{7.0}_{4.52}$ $\times 10^{9}$ & 0.32$^{0.51}_{0.21}$ & 0.12$^{0.21}_{0.07}$ & 0.39$^{0.61}_{0.26}$ & 2.80$^{3.56}_{2.29}$ & 9.37$^{12.2}_{6.81}$ $\times 10^{-4}$ & 3.3$^{5.3}_{1.9}$ $\times 10^{-4}$ \\
\\
\hline
\\
50\% & 2.2$^{2.7}_{1.8}$ & 0.83$^{1.02}_{0.68}$ & 1.11$^{1.36}_{0.91}$ $\times 10^{10}$ & 1.25$^{1.88}_{0.84}$ & 0.62$^{0.98}_{0.39}$ & 1.30$^{1.78}_{0.94}$ & 5.53$^{6.78}_{4.52}$ & 4.1$^{5.47}_{3.21}$ $\times 10^{-4}$ & 7.4$^{12}_{4.7}$ $\times 10^{-5}$ \\

\enddata

\tablenotetext{a}{Hydrogen number density in cm$^{-3}$.}
\tablenotetext{b}{Ionization timescale in cm$^{-3}$ s.}
\tablenotetext{c}{Emission measure (EM) in 10$^{58}$ cm$^{-3}$.}
\tablenotetext{d}{Swept gas mass in $M_\odot$.}
\tablenotetext{e}{Dust mass from fit to IRS spectra, in $M_\odot$.}

\tablecomments{Errors listed are 90\% confidence intervals from
$\chi^{2}$ fitting to the post-shock density in column 2. Porosity,
$\cal P$, is percentage of grain occupied by vacuum. Three rows for
compact grain case correspond to different values of shock compression
ratio of 2.65, 4.8, and 6.8, as described in text. LMC abundances are
assumed, so that $n_{e}$ = 1.2$n_{H}$. Line strengths are for shocked
ISM, and are reported as fractions of observed line strengths measured
from RGS spectra by model with absorption described in text. For this
remnant, observed line strengths are: O Ly$\alpha$ = 4.41 $\times
10^{-13}$ ergs cm$^{-2}$ s$^{-1}$; O K$\alpha$ = 5.05 $\times
10^{-13}$ ergs cm$^{-2}$ s$^{-1}$. Models predicting fractional line
fluxes of greater than 1 are ruled out. M$_{D}$ is total amount of
radiating dust, extrapolated from single slit position to entire
remnant. M$_{D}$/M$_{G}$ is ambient, {\it pre-shock} dust/gas ratio,
where sputtering effects are accounted for.}
\label{resultstable0509}
\end{deluxetable}

\begin{deluxetable}{lccccccccc}
\rotate
\vspace{-0.3truein}
\tablecolumns{10}
\tablewidth{0pc}
\tabletypesize{\footnotesize}
\tablecaption{Predicted ISM Oxygen Line Strengths for SNR B0519-69.0}
\tablehead{
\colhead{$\cal P$} & $n_{H}$\tablenotemark{a} & $n_{0}$\tablenotemark{a} & $n_{e}t$\tablenotemark{b} & EM\tablenotemark{c} & O Ly$\alpha$ & O K$\alpha$ & M$_{G}$\tablenotemark{d} & M$_{D}$\tablenotemark{e} & M$_{D}$/M$_{G}$ }

\startdata

0 & 6.2$^{6.8}_{5.6}$ & 2.34$^{2.57}_{2.11}$ & 4.69$^{5.14}_{4.24}$ $\times 10^{10}$ & 9.92$^{11.9}_{8.08}$ & 2.10$^{2.41}_{1.77}$ & 2.63$^{2.91}_{2.35}$ & 15.6$^{17.1}_{14.1}$ & 1.95$^{2.14}_{1.82}$ $\times 10^{-3}$ & 1.3$^{1.5}_{1.1}$ $\times 10^{-4}$ \\
0 & 6.2$^{6.8}_{5.6}$ & 1.29$^{1.42}_{1.17}$ & 4.69$^{5.14}_{4.24}$ $\times 10^{10}$ & 5.47$^{6.60}_{4.48}$ & 1.16$^{1.34}_{0.99}$ & 1.46$^{1.62}_{1.30}$ & 8.60$^{9.43}_{7.77}$ & 1.95$^{2.14}_{1.82}$ $\times 10^{-3}$ & 2.3$^{2.8}_{1.9}$ $\times 10^{-4}$ \\
0 & 6.2$^{6.8}_{5.6}$ & 0.91$^{1.0}_{0.82}$ & 4.69$^{5.14}_{4.24}$ $\times 10^{10}$ & 3.86$^{4.65}_{3.14}$ & 0.82$^{0.95}_{0.70}$ & 1.02$^{1.14}_{0.91}$ & 6.06$^{6.65}_{5.47}$ & 1.95$^{2.14}_{1.82}$ $\times 10^{-3}$ & 3.2$^{3.9}_{2.7}$ $\times 10^{-4}$ \\
\\
\hline
\\
25\% & 9.5$^{10.3}_{8.6}$ & 3.58$^{3.89}_{3.25}$ & 7.18$^{7.79}_{6.50}$ $\times 10^{10}$ & 23.2$^{27.4}_{1.91}$ & 3.83$^{4.29}_{3.38}$ & 4.19$^{4.60}_{3.79}$ & 23.9$^{25.9}_{21.6}$ & 9.34$^{10.2}_{8.74}$ $\times 10^{-4}$ & 3.9$^{4.7}_{3.4}$ $\times 10^{-5}$ \\
\\
\hline
\\
50\% & 17$^{18.5}_{15.6}$ & 6.42$^{6.98}_{5.89}$ & 1.29$^{1.40}_{1.18}$ $\times 10^{11}$ & 74.6$^{88.3}_{62.8}$ & 7.80$^{8.45}_{7.05}$ & 7.95$^{8.59}_{7.24}$ & 42.8$^{46.6}_{39.3}$ & 4.80$^{5.2}_{4.5}$ $\times 10^{-4}$ & 1.1$^{1.3}_{0.97}$ $\times 10^{-5}$ \\

\enddata

\tablenotetext{a}{Hydrogen number density in cm$^{-3}$.}
\tablenotetext{b}{Ionization timescale in cm$^{-3}$ s.}
\tablenotetext{c}{Emission measure (EM) in 10$^{58}$ cm$^{-3}$.}
\tablenotetext{d}{Swept gas mass in $M_\odot$.}
\tablenotetext{e}{Dust mass from fit to IRS spectra, in $M_\odot$.}

\tablecomments{Same as Table 2, except that measured line strengths
from RGS data are as follows: O Ly$\alpha$ = 5.81 $\times 10^{-13}$
ergs cm$^{-2}$ s$^{-1}$; O K$\alpha$ = 3.91 $\times 10^{-13}$ ergs
cm$^{-2}$ s$^{-1}$. Models predicting fractional line fluxes of
greater than 1 are ruled out.}
\label{resultstable0519}
\end{deluxetable}

\newpage
\clearpage

\begin{figure}
\includegraphics[width=15cm]{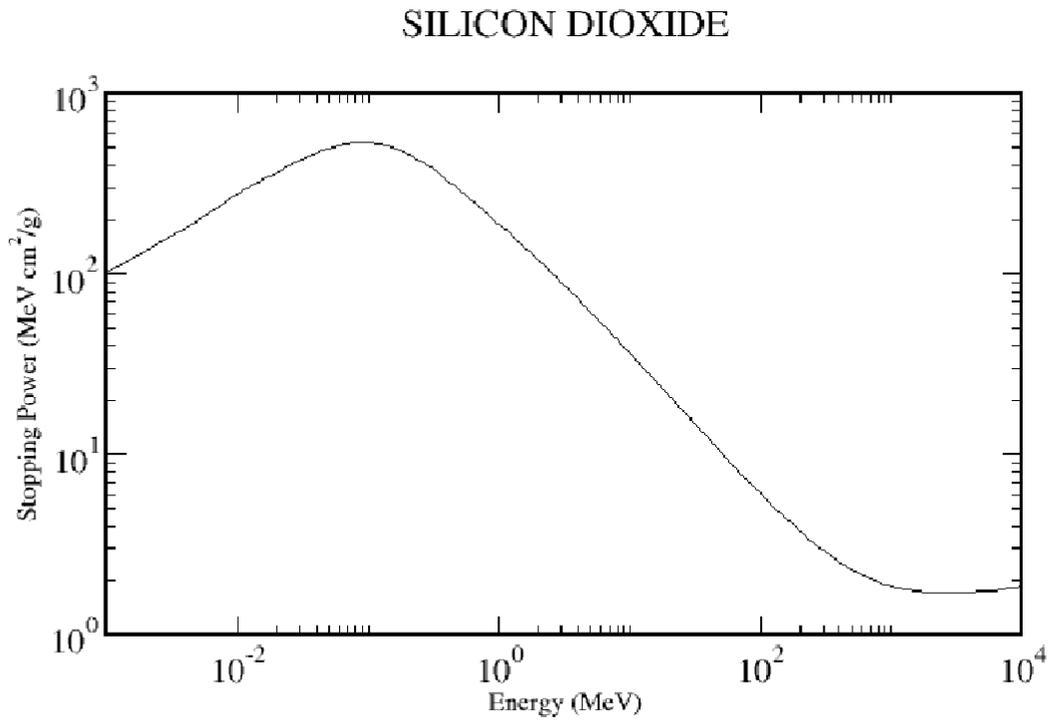}
\caption{Stopping power of a proton in silicon dioxide (SiO$_{2}$) as
a function of proton energy. Taken from the NIST {\it PSTAR} database
(see text).
\label{si02stopping}
}
\end{figure}

\begin{figure}
\includegraphics[width=16cm]{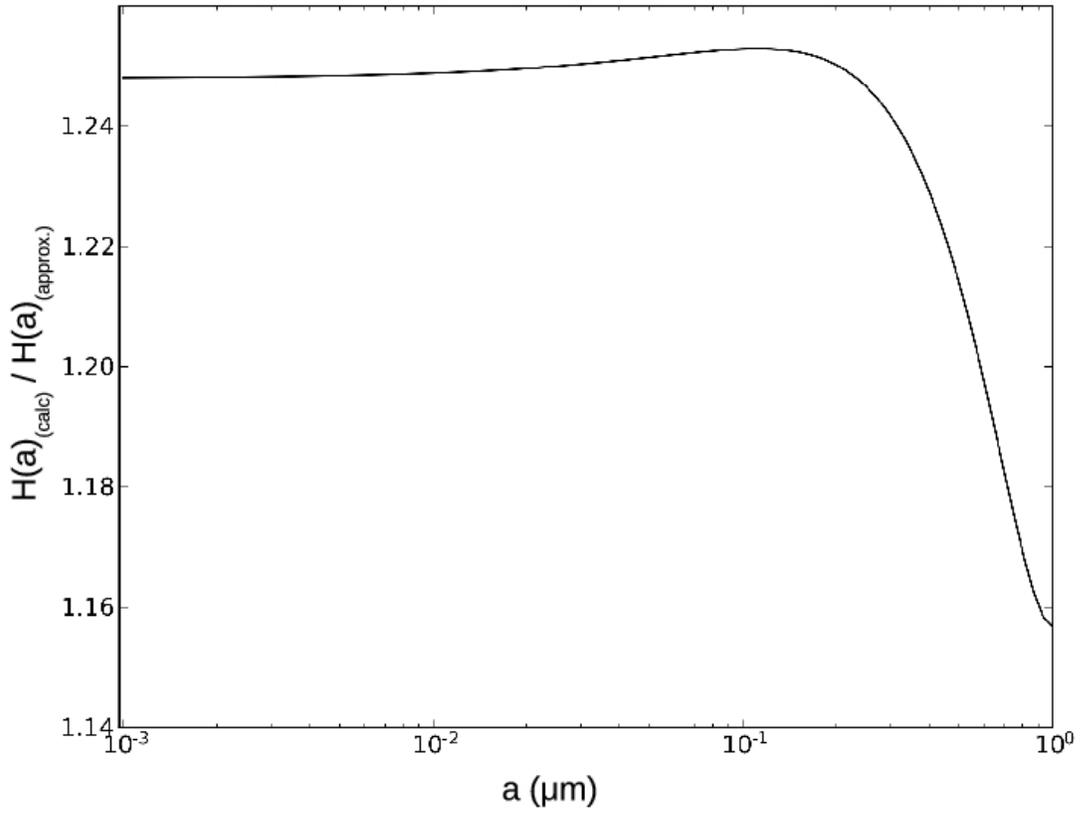}
\caption{Ratio of proton heating rate of a silicate grain from
calculations in this work to approximations from \citet{draine79} \&
\citet{dwek81} for an impinging particle energy of 100 keV, as a
function of grain radius.
\label{heatingrateratio}
}
\end{figure}

\begin{figure}
\includegraphics[width=16cm]{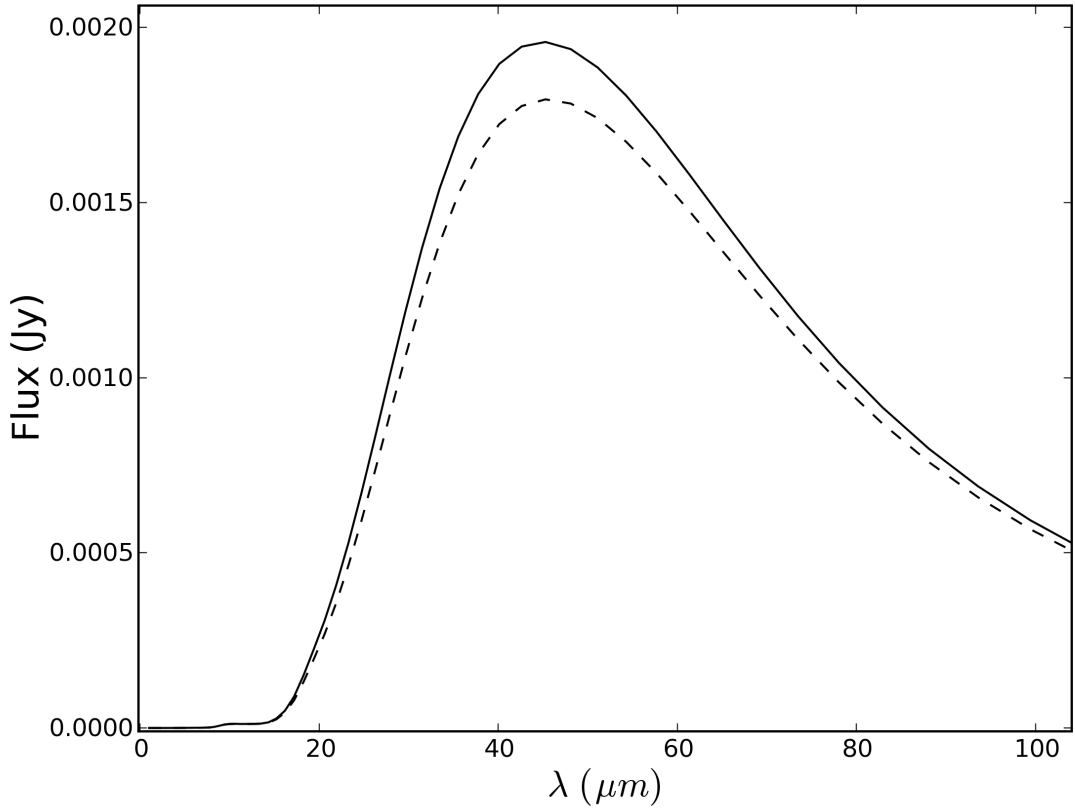}
\caption{Comparison of spectra produced in dust heating model for
protons of 100 keV. Solid line: spectrum assuming proton projected
ranges calculated in Section 7.5. Dashed line: spectrum assuming
analytical approximation to proton projected range from
\citet{draine79}. Electron heating ($T_{e}$ = 2 keV) is included in
this model. Projected range is defined as the average length traveled
by a particle into a material.
\label{spectracompare}
}
\end{figure}

\begin{figure}
\includegraphics[width=16cm]{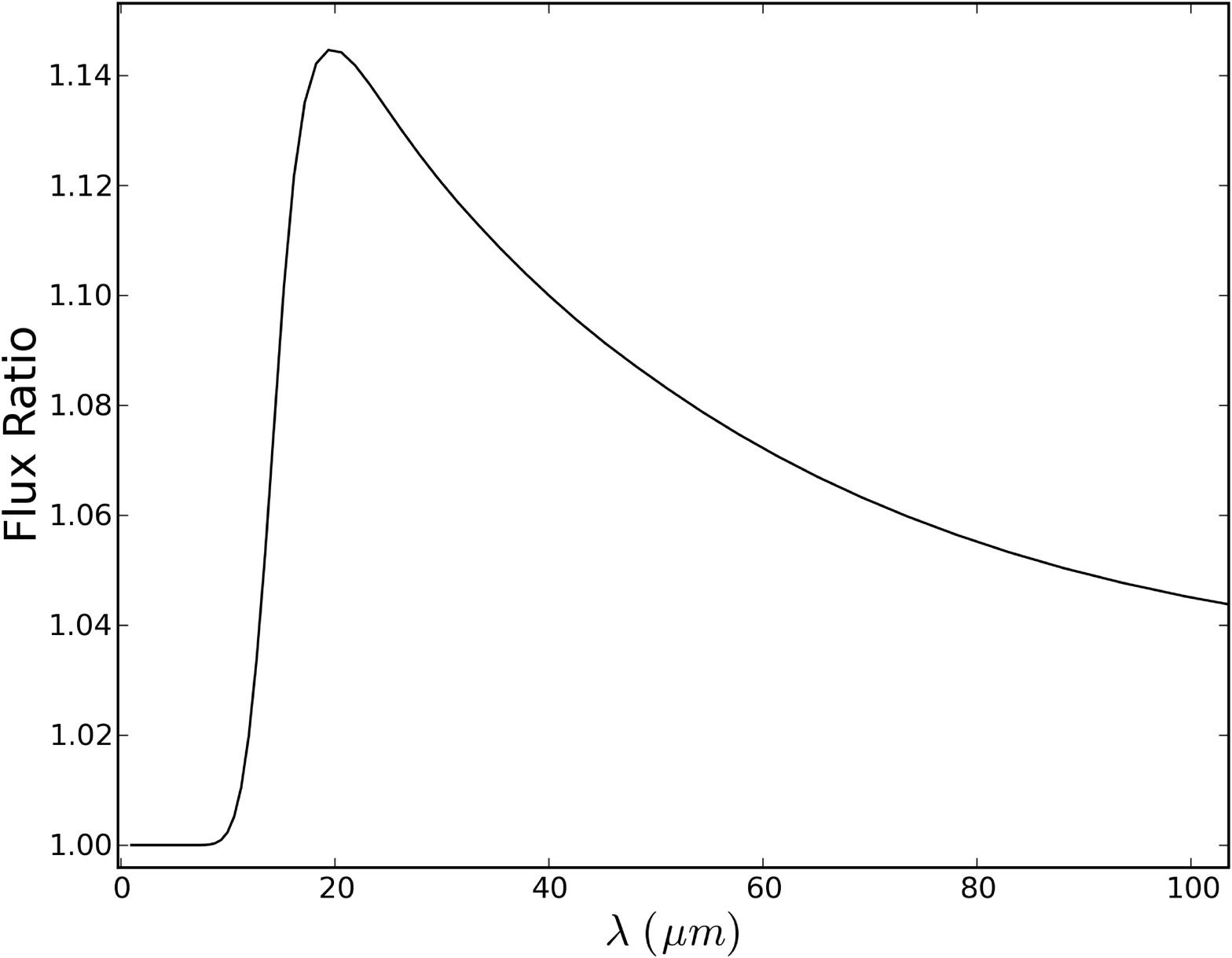}
\caption{Ratio of spectra from Figure~\ref{spectracompare}.
\label{spectraratio}
}
\end{figure}

\clearpage

\appendix
\section{Proton Energy Deposition Rates}

In young SNRs, and in 0509 in particular, proton temperatures are much
higher than electron temperatures, and heating by protons cannot be
neglected as it can in slower shocks where temperatures have had time
to equilibrate. Here we investigate whether simple analytic
approximations to proton and alpha particle energy deposition rates
are valid at ion energies approaching or exceeding 100 keV.

The stopping power (energy loss per unit path length) and projected
range (average length traveled by a particle into a material) for both
electrons and protons into a given grain are generally given by
analytic expressions approximating experimental data. For electrons,
these expressions fit the experimental data to within 15\% for the
energy range of 20 eV to 1 MeV \citep{dwek87}. For protons and alpha
particles, the approximate stopping power, based on experimental data
from \citet{andersen77} and \citet{ziegler77}, is given by
\citet{draine79} as

\begin{equation}
R_{H} = 3 \times 10^{-6} \rho^{-1} (E/keV) g\ cm^{-2}, R_{He} = 0.6R_{H},
\end{equation}

\noindent
where $\rho$ is the mass density of the grain. The authors caution
that this expression is valid only for $E < 100$ keV. We raise here
two questions: 1) Are these approximations appropriate for energies at
or exceeding the 100 keV threshold (as may well be the case in very
young SNRs with shock speeds exceeding 6000 km s$^{-1}$)? 2) These
approximations are based on data over 30 years old; are they still
valid?

To answer both of these questions, we turn to more recent laboratory
data on the stopping power of protons and alpha particles in various
materials, obtained from the online {\it PSTAR} and {\it ASTAR}
databases published and maintained by the National Institute of
Standards and Technology.\footnote{Databases are available at \tt
http://physics.nist.gov/PhysRefData/Star/Text/contents.html} In
Figure~\ref{si02stopping}, we show the stopping power of a proton in
silicon dioxide as a function of energy. The stopping power turns over
above $\sim 100$ keV. Since, for an extremely fast shock, there will
be a significant population of protons with energies in this range,
this turnover should be properly accounted for in calculating the
projected range of the particle.

Interstellar grains are more complicated than silicon dioxide
(although it is possible that SiO$_{2}$ is a minor component of ISM
dust). Since the NIST databases only have information available for
actual materials that can be measured in the lab, materials like
``astronomical silicate (MgFeSiO$_{4}$)'' are not available. The
individual stopping powers for the constituent elements are available,
however, and can be combined to approximate the stopping power for a
grain of arbitrary composition. To do this, we use ``Bragg's Rule,''
\citep{bragg05}, given by

\begin{equation}
S(A_{m}B_{n}) = m \cdot S(A) + n \cdot S(B),
\end{equation}

\noindent
where $S(A_{m}B_{n})$ is the total stopping power of a molecule
$A_{m}B_{n}$, and $S(A)$ and $S(B)$ are the stopping powers of the
individual constituents. Since Mg is not included in the NIST
database, we use the results of \citet{fischer96}. Thus, for
MgFeSiO$_{4}$, one has

\begin{equation}
S(MgFeSiO_{4}) = S(Mg) + S(Fe) + S(Si) + 4\cdot S(O).
\end{equation}

The calculated projected range for protons based on this stopping
power (similarly for graphite) are given by the following polynomial
expressions in logarithmic-space, where $E$ is the energy of the
impinging particle in keV:

\underline{Protons}

\begin{equation}
\log\ (R_{H,silicate}) = 0.053\ \log^{3}\ (E)-0.202\ \log^{2}\ (E)+1.21\ \log\ (E)-5.68
\end{equation}

\begin{equation}
\log\ (R_{H,graphite}) = 0.053\ \log^{3}\ (E)-0.158\ \log^{2}\ (E)+0.961\ \log\ (E)-5.46
\end{equation} 

\underline{Alpha Particles}

\begin{eqnarray*}
\log\ (R_{He,silicate}) &=& 3.78 \times 10^{-3}\ \log^{5}\ (E)-8.79 \times 10^{-3}\ \log^{4}\ (E)-4.25 \times 10^{-2}\ \\
& & \log^{3}\ (E)+6.25 \times 10^{-2}\ \log^{2}\ (E) + 1.04\ \log\ (E)-5.79
\end{eqnarray*}

\begin{eqnarray*}
\log\ (R_{He,graphite}) &=& 3.33 \times 10^{-3}\ \log^{5}\ (E)-6.13 \times 10^{-3}\ \log^{4}\ (E)-3.46 \times 10^{-2}\ \\
& & \log^{3}\ (E)+3.03 \times 10^{-2}\ \log^{2}\ (E) + 0.966\ \log\ (E) - 5.76.
\end{eqnarray*}

Ultimately, it is necessary to determine how much energy is deposited
into the grain by an impinging particle, in order to calculate grain
heating rates and determine whether these improved energy deposition
rates matter. In Figure~\ref{heatingrateratio}, we show the ratio of
the heating rate, $H(a,T)$, calculated from the NIST energy deposition
rates and that calculated from the analytical approximations of
\citet{draine79} and \citet{dwek81}, for a proton temperature of 100
keV, which is relevant for the fast shocks, such as those seen in
0509. Figure~\ref{spectracompare} shows a comparison of the spectra of
grains heated behind such a shock, for both the modified energy
deposition rates described above and the analytical
approximations. Figure~\ref{spectraratio} shows the ratio of the
spectra. The main difference is in the 18-40 $\mu$m range, but the
maximum difference is only $\sim 15$\%. The use of analytical
approximations to proton heating rates should be valid for most
cases. Nonetheless, we use the modified rates in reporting results in
this work.

\subsection{Moving Grains}

Because heating by protons is non-negligible in the case of such fast
shocks and high ion temperatures, it is necessary to account for
relative gas-grain motions between dust grains and ions in the
plasma. Grain heating rates for protons and alpha particles are
increased by 35 and 112\%, respectively, over the case where grains
are at rest with respect to the plasma. Grain heating is still
dominated by electrons, though, so the overall heating rate is
increased by only $\sim$ 25\% for the moving grain case. When folded
through the model, this results in a decrease of about 10\% in the
plasma densities necessary to fit IRS data. These numbers are reported
throughout the text. We assume that the charge-to-mass ratio of the
grains is sufficiently low that grains are not immediately affected by
the passage of the shock.

\end{document}